\documentclass{aa}  

\usepackage{graphicx}
\usepackage{txfonts}
\usepackage{amsmath}	
\usepackage{xspace}
\usepackage{xcolor}
\usepackage[colorlinks=true,citecolor=teal]{hyperref}

\newcommand{\msun}{$M_\odot$\xspace}

\newcommand{\g}[1]{\textcolor{black}{#1}}

\defcitealias{2024A&A...683A.139S}{SN24}

\begin{document}

   \title{On the uncertainty of the white dwarf astrophysical gravitational wave background}

\titlerunning{The WD astrophysical GW background}

   \author{Sophie Hofman
          \inst{1}\fnmsep\thanks{E-mail: shofman.astro@gmail.com},
          \and
          Gijs Nelemans\inst{1, 2, 3}
          }

\authorrunning{S. Hofman \& G. Nelemans}

   \institute{Department of Astrophysics/IMAPP, Radboud University, PO Box 9010,
6500 GL, The Netherlands
         \and
            Institute of Astronomy, KU Leuven, Celestijnenlaan 200D, 3001 Leuven, Belgium
        \and
            SRON, Netherlands Institute for Space Research, Niels Bohrweg 4, 2333 CA Leiden, The Netherlands
             }

   \date{Received \today; accepted }

 
  \abstract
   {The astrophysical gravitational wave background (AGWB) is a stochastic gravitational wave (GW) signal that is emitted by different populations of inspiralling binary systems containing compact objects \g{throughout the Universe}. In the frequency range between 10$^{-4}$ and 10$^{-1}$ Hz it will be detected by future space-based gravitational wave detectors like the Laser Interferometer Space Antenna (LISA). Recently, we concluded that the white dwarf (WD) contribution to the AGWB dominates over that of black holes (BHs) \g{and neutron stars (NSs)}.}
   { We aim to investigate the uncertainties of the WD AGWB that arise from the use of different stellar metallicities, different star formation rate density (SFRD) models, and different binary evolution models.}
   {We use the code developed before to determine the WD component of the AGWB. We use a metallicity dependent SFRD based on earlier work to construct five different SFRD models. We use four different population models that use different common-envelope treatment and six different metallicities for each model.}
   {For all possible combinations, the WD component of the AGWB is dominant over other populations of compact objects. The effects of metallicity and population model are smaller than the effect of a (metallicity dependent) SFRD model. We find a range of about a factor of 5 in the level of the WD AGWB around a mid value of $\Omega_{\rm WD} = 4\times 10^{-12}$ at 1 mHz and a shape that depends weakly on the model.
   }
   {We find an uncertainty for the WD component of the AGWB of about a factor 5. We note that there exist other uncertainties that have an effect on this signal as well. We discuss whether the turnover of the WD AGWB at 10 mHz will be detectable by LISA, and find that this is likely. We confirm the previous finding that the WD component of the AGWB dominates over other populations, in particular BHs.  }


   \maketitle
%

\section{Introduction}

In the visible Universe, there are a very large number of gravitational wave (GW) sources that consist of two stellar mass compact objects, white dwarfs (WDs), neutron stars (NSs) or black holes (BHs). With a GW detector typically only a (tiny) selection of these sources can be individually detected, and many cannot. The Laser Interferometer Space Antenna (LISA) mission \citep{2017arXiv170200786A}, for instance can individually detect a large number of binaries (mostly double WD) in the Milky Way \citep[e.g.][]{1987A&A...176L...1L,1987ApJ...323..129E,1990ApJ...360...75H,2001A&A...375..890N,2010ApJ...717.1006R,2013MNRAS.429.1602Y,2019MNRAS.490.5888L,2022MNRAS.511.5936K,2023A&A...669A..45T}. \g{The remaining (millions) of binaries in the Milky Way form} a collective stochastic foreground of Galactic sources that is highly anisotropic \citep[e.g.][]{2012ApJ...758..131N,2021PhRvD.104d3019K}. \g{However, there is a huge number of unresolved compact binary sources in the extragalactic Universe that form } a stochastic isotropic signal \citep[e.g.][]{2003MNRAS.346.1197F}. This Astrophysical GW Background (AGWB) \textcolor{black}{, the background signal in GW solely made up of astrophysical sources,} encodes the combined information about the different source populations \citep[see, e.g.][]{2001MNRAS.324..797S,2003MNRAS.346.1197F,2015A&A...574A..58K, 2023LRR....26....2A}. 
It is a broadband signal \citep{2001astro.ph..8028P} that extends from the high-frequency band covered by the LIGO/Virgo/KAGRA detectors \citep{2015CQGra..32g4001L,2015CQGra..32b4001A,2013PhRvD..88d3007A} to the mHz band covered by future space GW detectors such as LISA \citep{2017arXiv170200786A}, Taiji \citep{2021PTEP.2021eA108L} or TianQin \citep{2016CQGra..33c5010L}. \g{Its amplitude gives an estimate of the total number of sources, weighted with their redshift distribution, while its shape potentially encodes details of the population, something we study in more detail in this paper. Since it is not likely that in particular the WD populations can be probed in any other way, this provides a unique probe of binary evolution at high redshift.}

The AGWB is not only interesting in itself as probe of the large population of binaries across cosmic time that cannot be detected individually, but also competes with potential other stochastic GW backgrounds that have an origin in the Early Universe and probe fundamental physics \citep[see e.g.][]{2009JCAP...12..024C,2012JCAP...06..027B,2023LRR....26....5A}. \textcolor{black}{While LISA will be able to detect the AGWB, one of the main objectives is to be able to detect the cosmological GWB (caused by e.g. cosmic strings \citep{Caprini_2018, 2023LRR....26....5A}). Understanding of the AGWB will therefore be of considerable importance to achieve the LISA objective to investigate the cosmological GWB \citep{2023MNRAS.526.4378L}}.

In the LISA band, the AGWB due to BH mergers could be detected \citep[e.g.][]{2016MNRAS.461.3877D,2022A&A...660A..26B,2023JCAP...08..034B,2023MNRAS.526.4378L}, but \citet{2024A&A...683A.139S} recently showed that the AGWB due to WD binaries \citep[studied earlier by][]{2003MNRAS.346.1197F} likely dominates \textcolor{black}{over that due to BHs}. \g{The AGWB due to NSs is expected to be significantly below even that of BHs because of the relative small number of NS mergers compared to WD mergers and their relative low mass compared to BHs \citep[e.g.][]{2023JCAP...08..034B,2023PhRvX..13a1048A}}. However, \citet{2024A&A...683A.139S} only considered a single model for the double WD population in the Universe and a simplified model for the star formation history. \g{Yet, we know that the metallicity of the progenitor stars may have a significant impact on the compact binary formation \cite[e.g.][]{2018MNRAS.480.2011G,2018MNRAS.474.2937C,2019MNRAS.490.3740N} and the metallicity of star formation is different at different times in the Universe. In addition, the star formation history of the Universe is uncertain \citep[e.g.][]{2019MNRAS.488.5300C}}. Here, we investigate the influence of the assumed star formation history of the Universe and the metallicity distribution of that star formation as well as some different binary evolution models on the WD AGWB. 

The paper is organised as follows. In Sect.~\ref{methods} we present the model we use for the calculation of the AGWB as well as the different assumptions about binary evolution and star formation histories. In Sect.~\ref{results} we present the results of our calculations and we conclude with a discussion of the limitations of calculation, possible future work and the main conclusions (Sect.~\ref{discussion}).

\section{Methods}\label{methods}

\subsection{Calculating the WD AGWB}

We use the code developed by \citet{2024A&A...683A.139S} (hereafter \citetalias{2024A&A...683A.139S}) to calculate the AGWB. Schematically, the AGWB in a given frequency bin ($f_r$) is computed by adding up the contributions from all WD binaries that emit GW with frequencies ($f_e$) that are redshifted into that bin ($f_e = f_r (1 + z)$) in 20 redshift bins between z = 0 and z = 8. The population of WD binaries in each redshift bin is obtained by integrating the star formation in the relevant volume that happened before the cosmic time associated with that redshift and selecting the WD binaries that emit at the right frequencies. Following \citet{2003MNRAS.346.1197F}, for each bin, the AGWB is calculated in terms of the dimensionless energy density spectrum 
\begin{equation}\label{eq:Omega_Fr}
    \Omega(f_r) \approx \frac{1}{\rho_c c^3} \frac{f_r F_{f_{r_1}\to f_{r_2}}}{f_{r_2} - f_{r_1}}\,,
\end{equation}
where $F_{f_{r_1} \to f_{r_2}}$ is the GW flux received from the sum of the populations of sources in the different redshift bins. See \citetalias{2024A&A...683A.139S} for more details of the calculation. 

In order to calculate the GW flux, the population of WD binaries emitting GWs in a particular band in a redshift bin needs to be simulated. To do so, two crucial ingredients are needed: a model for how binaries evolve from their formation as main sequence stars to the stage where they emit GWs plus a model for how many binaries were formed in the history of the Universe, the star formation rate density (SFRD, i.e. the amount of star formation per unit time per unit volume) over cosmic time. In \citetalias{2024A&A...683A.139S}, the SFRD was approximated by a single function as given by \citet{madau_cosmic_2014}. For the binary model, a single model at Solar metallicity, based on the SeBa population synthesis code \citep{1996A&A...309..179P,2001A&A...365..491N,2012A&A...546A..70T} was used. In the next sections we will introduce the variations we use in this study.

\subsection{The star formation rate density}

\begin{figure}
    \includegraphics[width=\columnwidth]{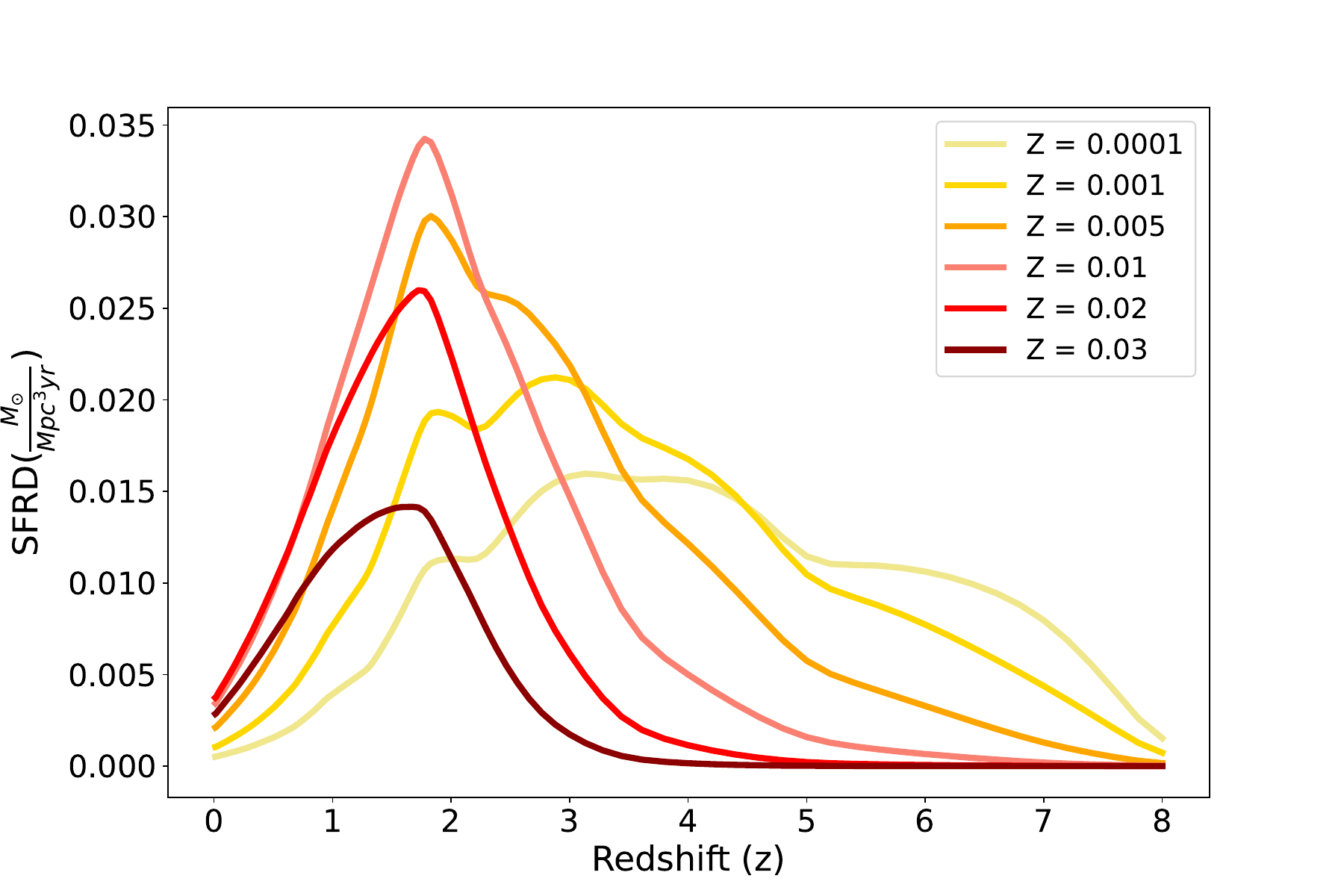}
    \caption{Total star formation rate density (SFRD) in $M_\odot$ Mpc$^{-3}$ yr$^{-1}$ versus redshift (z), for the six metallicity bins for the MZ19 SFRD model.}
    \label{fig:SFRD vs z per bin MZ19}
\end{figure}

\begin{figure}
    \includegraphics[width=\columnwidth]{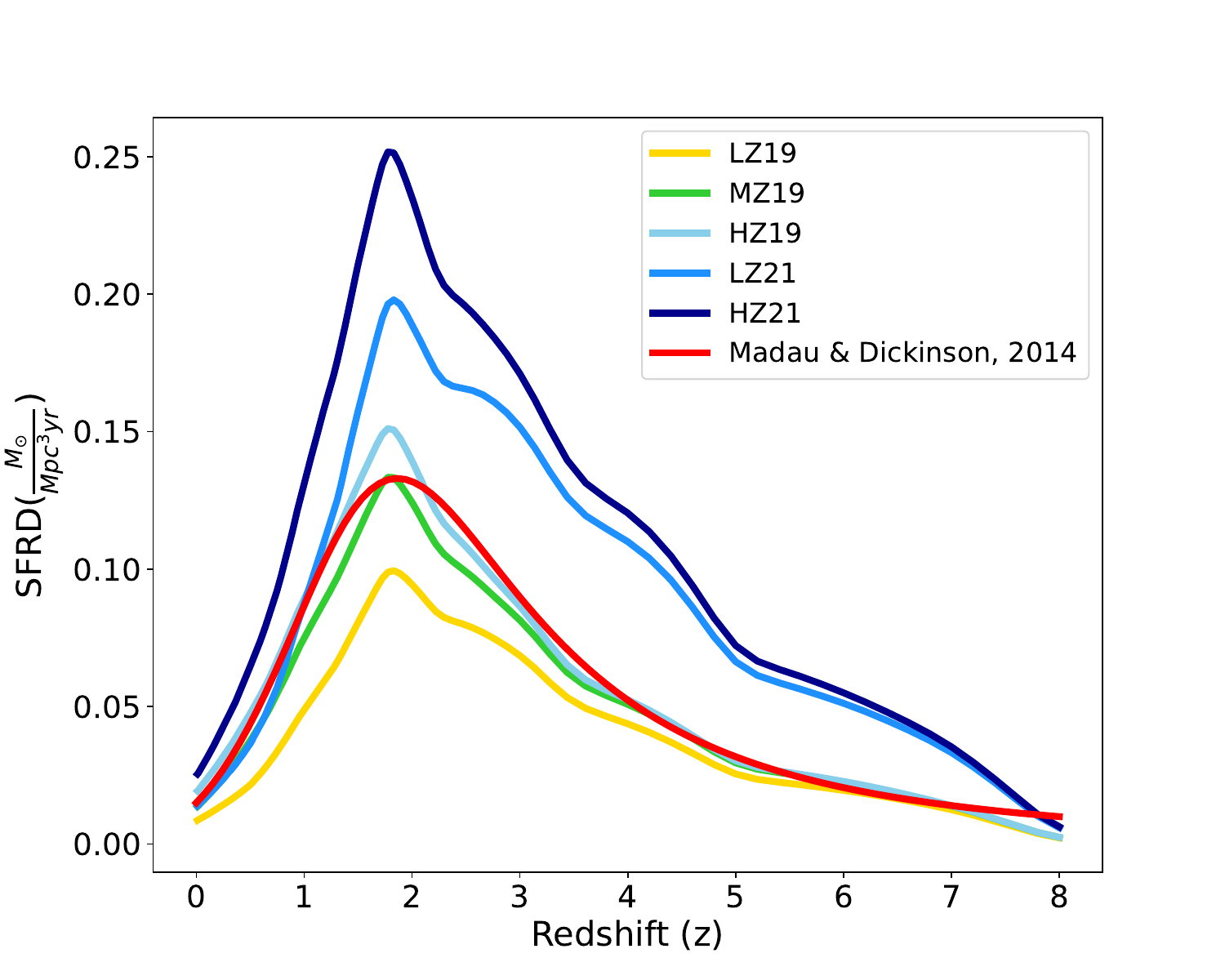}
    \caption{Total star formation rate density (SFRD) in $M_\odot$ Mpc$^{-3}$ yr$^{-1}$ versus redshift (z), for the six different SFRD models. \textcolor{black}{The red line represents the \citet{madau_cosmic_2014} model used in \citetalias{2024A&A...683A.139S}. The other five lines represent five of the SFRD models introduced in \citet{2019MNRAS.488.5300C,2020A&A...636A..10C,2021MNRAS.508.4994C}. The bottom three models are the low- intermediate- and high-metallicity extremes without starbursts and the top two the extreme models including starbursts (see text for more details).}}
    \label{fig:SFRD vs z}
\end{figure}

We aim to include a metallicity dependent SFRD in order to make the calculation more realistic, but also to study the impact of metallicity on the AGWB (and thus the potential of using a measurement of the AWGB to learn about the metallicity dependence of the SFRD). We base our calculations on the series of papers \citet{2019MNRAS.488.5300C,2020A&A...636A..10C,2021MNRAS.508.4994C} in which observational constraints on the number of galaxies, their star formation rate and their metallicity are combined. From these, several variations of possible SFRDs are constructed: a standard set of variations, based on the 2019 paper, with a intermediate-metallicity model (MZ19), plus a low-metallicity extreme (LZ19) and a high-metallicity extreme (HZ19). \citet{2021MNRAS.508.4994C} add the effect of star burst galaxies, leading to two new variations, one low-metallicity extreme (LZ21) and a high-metallicity extreme (HZ21). 

With these models the metallicity evolution of the star formation can be followed. As an example we show in Fig.~\ref{fig:SFRD vs z per bin MZ19} the SRFD of the MZ19 model, split in six different metallicity bins, with bin centers ranging from Z = 0.0001 to Z = 0.03. The exact ranges are [0, 0.0005], (0.0005, 0.003], (0.003, 0.008], (0.008, 0.015], (0.015, 0.03], ($>$0.03). As expected, at high redshift the star formation is dominated by low metallicities, while in the lower redshift range the higher metallicities dominate. The SFRD split into the six metallicity bins for the LZ21 and HZ21 model are shown in Appendix \ref{appendix} (the LZ19 and HZ19 plots are very similar but a lower total due to the lack of starburst galaxies).

Apart from the distribution over metallicity, the different variations, that sample some of the most important uncertainties in the observational constraints as discussed in the series of papers, also vary in the total SFRD over cosmic time. That directly translates in a different total numbers of stars formed. In Fig.~\ref{fig:SFRD vs z} we show the total SFRD for the five models mentioned above plus the \citet{madau_cosmic_2014} model used in \citetalias{2024A&A...683A.139S}. It is clear that the low-metallicity models also have lower total SFRD and that the addition of starbursts adds a significant amount of star formation leading to SFRDs that can be a factor of 2 larger than those without starbursts.

\subsection{Binary population models}

\begin{figure*}
\sidecaption
    \centering \includegraphics[width=12cm]{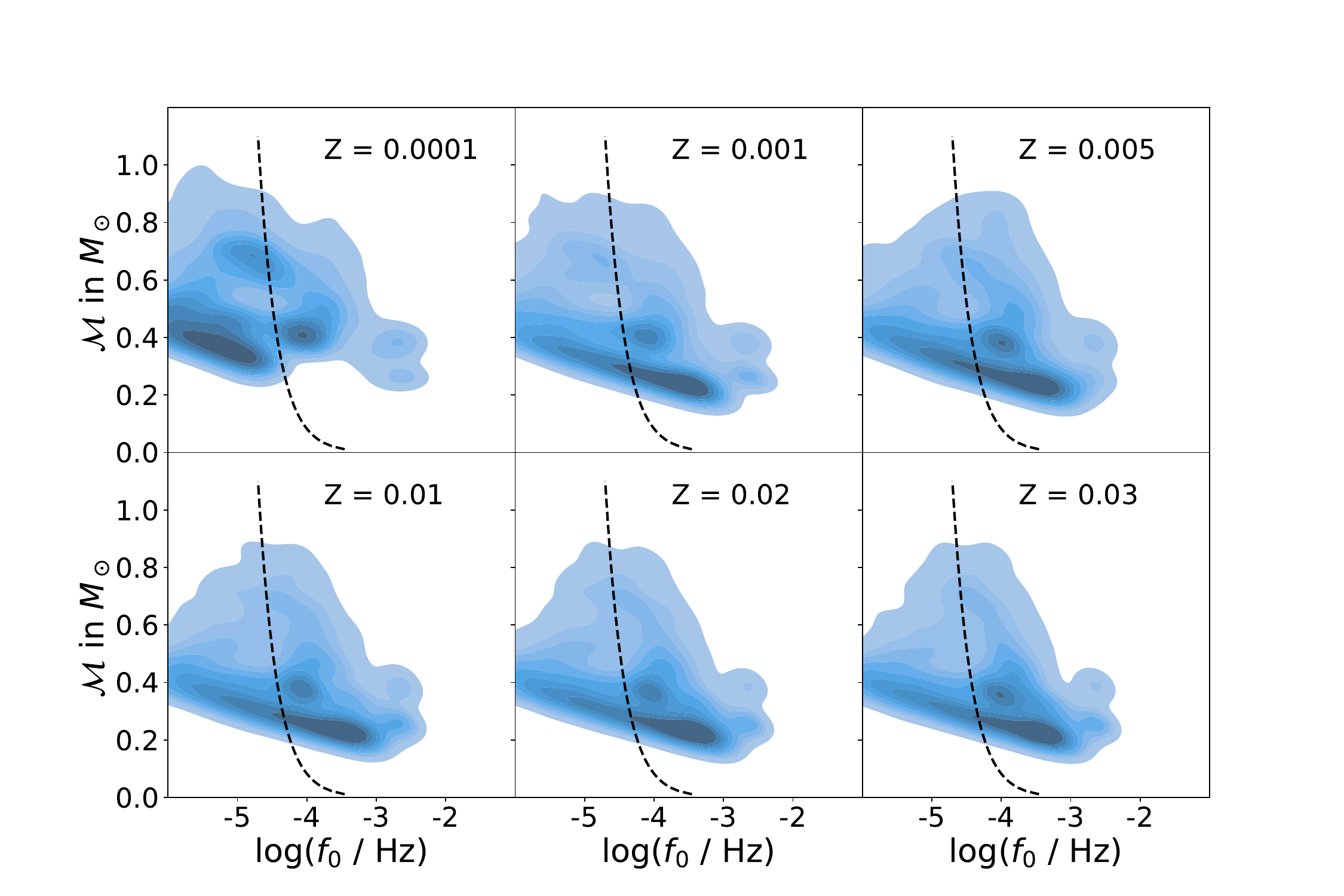}
    \caption{Density plot of the initial properties of the WD population in the case of a $\gamma\alpha$, $\alpha$ = 4 population model: chirp mass, $\mathcal{M}$, versus GW frequency at the time of formation. Each panel shows a different metallicity of the Universe. The dashed lines indicate frequencies above which there is significant (10\%) frequency evolution in a Hubble time.}
    \label{fig:Density plots ga4}
\end{figure*}

Now that we use metallicity dependent SFRDs we also need binary population models that are metallicity dependent. In addition, we study four different binary populations, that differ in the treatment of the common-envelope phase, that plays an important role in producing close double WDs. 

All models are calculated with the the SeBa population synthesis code \citep{1996A&A...309..179P,2001A&A...365..491N,2012A&A...546A..70T} that was also used by \citetalias{2024A&A...683A.139S} and to make synthetic models for the population of double WDs detectable by LISA in the Galaxy and nearby galaxies \citep{2001A&A...375..890N,2020A&A...638A.153K,2023MNRAS.521.1088K} that are used in the LISA Data Challenges \citep{2022arXiv220412142B}. 

For each of the binary models we run the code for six different metallicities, to cover the bins in which we split up the SRFD above: Z = 0.0001, 0.001, 0.005, 0.01, 0.02, 0.03. For each run we simulate 250,000 binaries with primary masses between 0.8 and 11 \msun, distributed according to the \citet{2001MNRAS.322..231K} initial mass function, with a flat mass ratio distribution and initial separations flat in $\log a$. Taking into account the whole mass range (i.e. also those not simulated in detail) and a mass-dependent binary fraction \citep[based on][]{2017ApJS..230...15M,2013A&A...552A..69V}, this corresponds to a total mass of star formation of $3.4 \times 10^6$ \msun.

For the formation of double WDs, one of the most important uncertainties is the effect of the two mass-transfer phases on the orbit. Because the stars typically start mass transfer when the donor has a deep convective envelope, the mass transfer is assumed to be unstable \citep[e.g. \citealt{1998MNRAS.296.1019H} but see][]{2023A&A...669A..45T}. Earlier studies have suggested that the standard energy-balance ($\alpha$) common-envelope cannot be a good description of the first phase of mass transfer \g{since it inevitably leads to double WDs in which the last formed WD has a lower mass, which is not observed}, but that a criterion based on angular momentum balance ($\gamma$) fits better \citep[see][]{2012A&A...546A..70T,2005MNRAS.356..753N,2001A&A...365..491N}. \g{Some works have suggested the first mass transfer is in fact stable \citep{2012ApJ...744...12W,2023A&A...669A..82L}, but that assumption leads to the opposite situation in which the last formed WD is the more massive one, which is also not always the case.} In the classical common-envelope description, $\alpha$ nominally represents the efficiency with which orbital energy is used to unbind the envelope, i.e. with a strict upper bound on $\alpha$ of 1. However, the common-envelope phase is complicated and additional energy sources may be available, potentially leading to higher values. Indeed reconstruction of double WD evolution, suggests values as high as 4 \citep[assuming a structure parameter $\lambda = 0.5$, see e.g.][]{2005MNRAS.356..753N}.

We run two models $\alpha \alpha$, assuming the standard common envelope in all cases with a structure parameter $\lambda = 0.5$, one with $\alpha =1$ and one with $\alpha = 4$. We also run two models $\gamma \alpha$, assuming the first mass transfer is described by $\gamma = 1.75$ and the second common envelope either by $\alpha =1$ or by $\alpha = 4$. The latter we consider our standard model. 

In Fig.~\ref{fig:Density plots ga4} we show the initial distribution of double WDs for this standard model as a function of initial GW frequency and chirp mass for the six different metallicities. Apart from the lowest metallicity, the bulk of the population is concentrated at low chirp mass and frequencies just above the line that indicates significant orbital evolution in a Hubble time. \g{The lowest metallicity looks different, because at low metallicity the stars are more compact leading to more massive cores when stars interact. This leads to higher chirp masses, but more importantly, after the first mass transfer, to tighter orbits that in the second mass transfer lead to merger of the system or, in case the more compact stars makes that the first mass transfer avoids the first giant branch, to wider systems that lead to wider final orbits}. The distribution changes only slowly with metallicity. For the other models the distributions are shown in Appendix \ref{appendix}. They all have similar average chirp masses around 0.45 $M_\odot$ with slightly higher values for the lowest metallicities.


\section{Results}\label{results}
The figures shown in this section all show the population model $\gamma\alpha$ with $\alpha$ = 4, unless stated otherwise.

%
\subsection{The effect of metallicity}

\begin{figure}
    \includegraphics[width=\columnwidth]{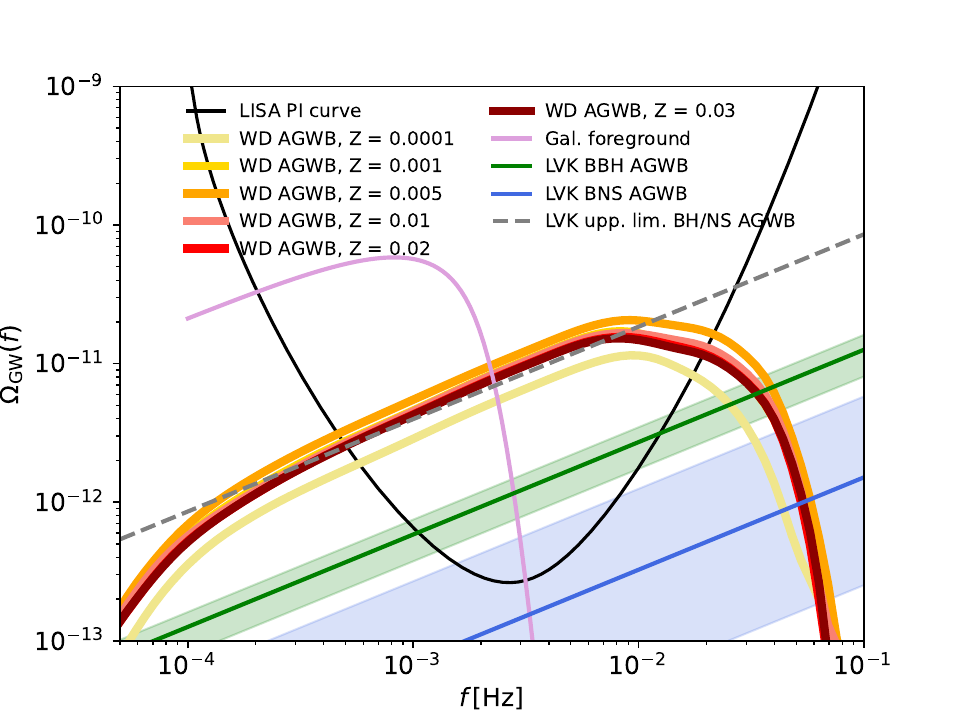}
    \caption{WD components of the AGWB for six different metallicities, compared to the LVK results (upper limit to BH/NS AGWB \citep[dashed grey][]{2021PhRvD.104b2004A} and estimates for the BBH and BNS components in green and blue, \citealt{2023PhRvX..13a1048A}), the LISA Powerlaw Integrated sensitivity \cite[black][]{2013PhRvD..88l4032T,2020PhRvD.101l4048A}, and an estimate of the Galactic foreground (pink) based on \cite{2021PhRvD.104d3019K}. The population synthesis model used is $\gamma\alpha$, $\alpha$ = 4 and the SFRD model used is that of \cite{madau_cosmic_2014}.}
    \label{fig:Omega ga4 SFH madau&dickinson}
\end{figure}

We first look at the influence of metallicity on the WD component of the AGWB in an artificial way, by simulating the AGWB assuming all star formation is at a particular metallicity. The result is shown in Fig.~\ref{fig:Omega ga4 SFH madau&dickinson}, where the green and the blue line represent the estimate for the \textcolor{black}{binary black hole (BBH) and binary neutron star (BNS) component (see \citealt{2023PhRvX..13a1048A})} and the dark red to yellow signals represent the WD component of the AGWB for the six different metallicities, respectively. The SFRD introduced in \citet{madau_cosmic_2014} is used for each of the metallicities. It is clear to see that a metallicity of 0.0001 leads to the lowest signal, whereas a metallicity of 0.005 leads to the highest signal. The other four metallicities are indistinguishable. The difference between the various signals, coming from the use of different metallicities, is approximately a factor of 1.9 at a frequency of 1 mHz. However, all six different metallicities lead to a WD component of the AGWB that is clearly dominant over the BBH and BNS components of the AGWB. Fig.~\ref{fig:Omega ga4 SFH madau&dickinson} shows the results for a population model of $\gamma\alpha$ with $\alpha$ = 4. The three other population models are shown in Appendix \ref{appendix}.

\subsection{The effect of different SFRD models}

\begin{figure}
    \includegraphics[width=\columnwidth]{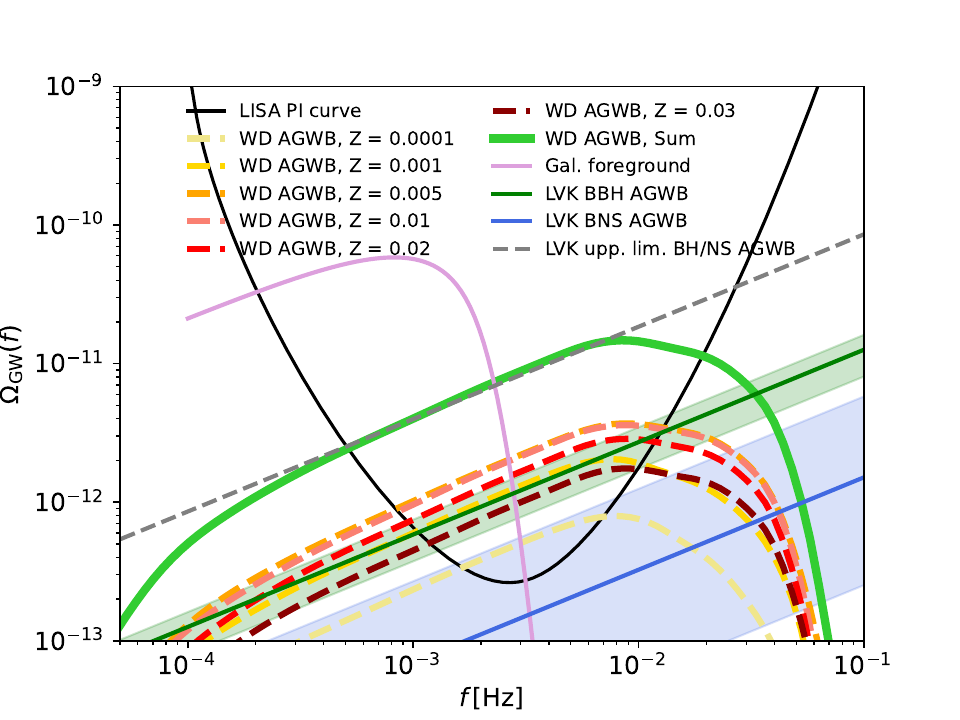}
    \caption{As in Fig.~\ref{fig:Omega ga4 SFH madau&dickinson}. The MZ19 SFRD model is used. The dashed lines are the WD components of the AGWB for each of the six metallicity bins, the solid light green line is the sum of all six separate WD components.}
    \label{fig:Omega ga4 MZ19}
\end{figure}

\begin{figure}
    \includegraphics[width=\columnwidth]{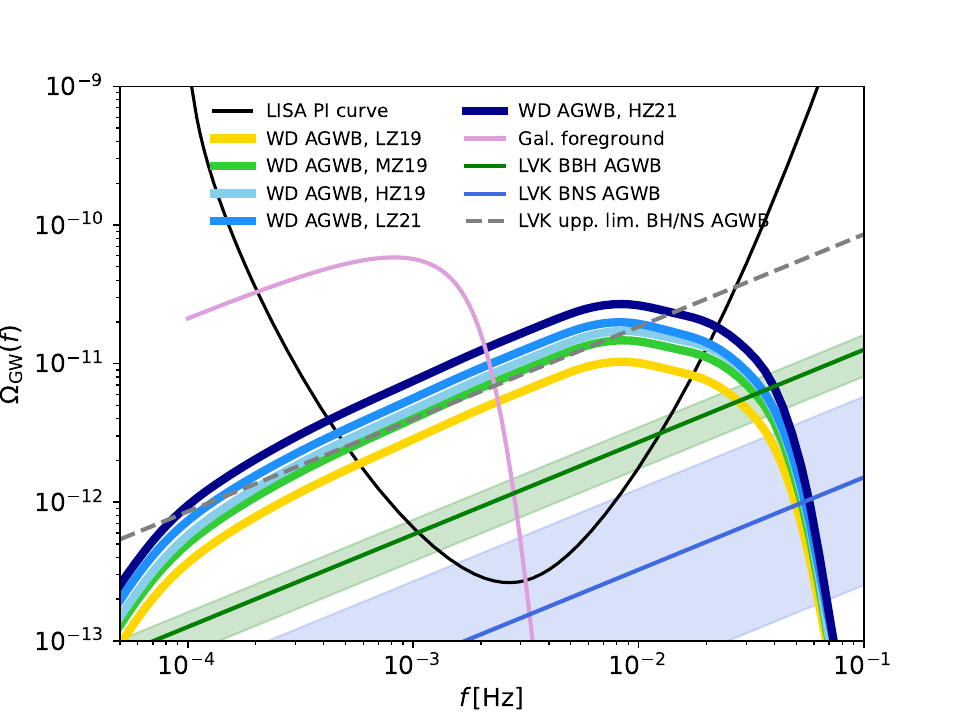}
    \caption{As in Fig.~\ref{fig:Omega ga4 SFH madau&dickinson}. Each line represents the sum of the signal for a different SFRD model. The light green line is the same as the one in Fig.~\ref{fig:Omega ga4 MZ19}.}
    \label{fig:Omega ga4 different SFRDs}
\end{figure}

In reality, the star formation is distributed over different metallicities. We show the effect of different SFRD models in Fig.~\ref{fig:Omega ga4 MZ19} and Fig.~\ref{fig:Omega ga4 different SFRDs}, where the standard components are as in Fig.~\ref{fig:Omega ga4 SFH madau&dickinson}. In Fig.~\ref{fig:Omega ga4 MZ19}, the dark red to yellow signals again represent the WD component of the AGWB for the six different metallicities, only this time, the MZ19 SFRD model by \citet{2019MNRAS.488.5300C} is used. Moreover, the SFRD is binned in the six metallicity bins instead of using the total SFRD, as done in Fig.~\ref{fig:Omega ga4 SFH madau&dickinson}. The light green line represents the sum of all six WD components of the AGWB. Fig.~\ref{fig:Omega ga4 MZ19} shows that the highest and the lowest metallicity lead to the weakest AGWB signal. This is as expected, since these metallicities have lower total SFRDs in their respective bins, as seen in Fig.~\ref{fig:SFRD vs z per bin MZ19}. Similarly, the metallicities leading to the highest AGWB signals, Z = 0.005, Z = 0.01 and Z = 0.02, also relate to the bins that encompass the highest total SFRDs as, again, seen in Fig.~\ref{fig:SFRD vs z per bin MZ19}. 

We plot the light green line from Fig.~\ref{fig:Omega ga4 MZ19} again in Fig.~\ref{fig:Omega ga4 different SFRDs} to illustrate the effect of the usage of different SFRD models. The five different SFRD models are clearly ordered in the same way as in Fig.~\ref{fig:SFRD vs z}, where the inclusion of starburst galaxies leads to higher AGWB signals than when starburst galaxies are omitted. Again, it is evident that the components of the AGWB due to double WDs, lead to higher signals than the AGWB due to BBHs and BNSs.

The usage of these five different SFRD models leads to a factor difference in the strength of the AGWB signal of about 2.6 at 1 mHz. This suggests that using different SFRD models has a larger effect on the strength of the AGWB signal compared to using different metallicities.

\subsection{The effect of different population models}

\begin{figure}
    \includegraphics[width=\columnwidth]{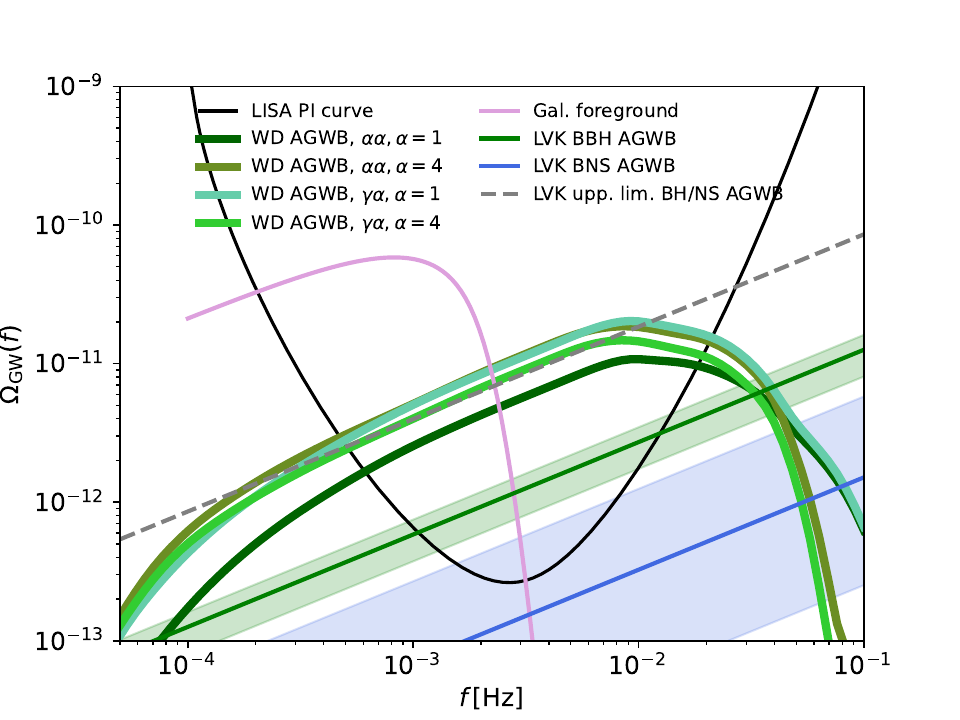}
    \caption{As in Fig.~\ref{fig:Omega ga4 SFH madau&dickinson}. Each line represents a different choice for population synthesis model. The MZ19 SFRD model is used.}
    \label{fig:Omega pop synth comparison}
\end{figure}

After investigating the effect of both metallicity and SFRD model, we also look at the effect of using a different population model. Up until now, the results were only shown for one single population model, namely $\gamma\alpha$, with $\alpha$ = 4. Fig.~\ref{fig:Omega pop synth comparison}, where the standard components are again as in Fig.~\ref{fig:Omega ga4 SFH madau&dickinson}, shows the effect of the usage of different population models. The light green signal as shown in previous figures, represents the previously used model, $\gamma\alpha$, with $\alpha$ = 4. The other three signals represent each of the other three population models, respectively.

It is clear to see that a population model with $\alpha\alpha$, with $\alpha$ = 1, leads to the lowest AGWB signal due to merging double WDs. This can be explained by the fact that a lower value for $\alpha$ leads to a tighter orbit. This means that there are more systems in this model that merge before the second common-envelope phase occurs, which explains the low AGWB signal compared to the other three population models. The other three signals are almost indistinguishable from each other, with the model of $\gamma\alpha$, with $\alpha$ = 4, leading to a slightly lower signal than the other two models. There is a factor difference of roughly 2 between the highest and lowest AGWB signal at a frequency of 1 mHz, coming from using a different population model. However, it should be noted that previous research (see Sect.~\ref{methods}) has shown that $\gamma\alpha$, with $\alpha$ = 4, shows the most agreement with observations of double WDs.

\subsection{Estimation of the AGWB}

\begin{figure*}
 \sidecaption   \includegraphics[width=1.4\columnwidth]{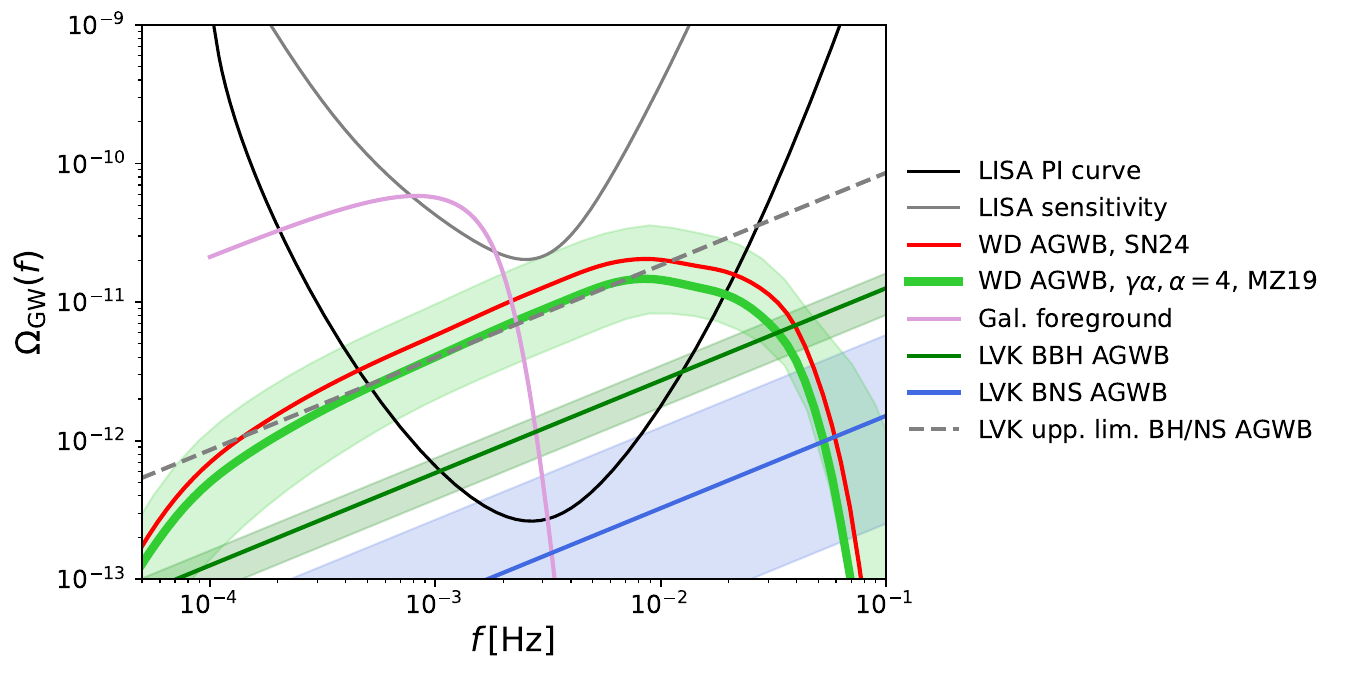}
    \caption{As in Fig.~\ref{fig:Omega ga4 SFH madau&dickinson}. The red line represents the WD component of the AGWB for a choice of $\alpha\alpha$, $\alpha$ = 4, with Z = 0.02 and the SFRD of \cite{madau_cosmic_2014}, which is the result from \citetalias{2024A&A...683A.139S}. The light green line represents the WD component of the AGWB for a choice of $\gamma\alpha$, $\alpha$ = 4 with the MZ19 SFRD model. The light green band represents the uncertainty estimate. The solid grey line shows the normal LISA sensitivity (i.e. not integrated over time and frequency).}
    \label{fig:Omega final result}
\end{figure*}

\begin{figure}
	\includegraphics[width=\columnwidth]{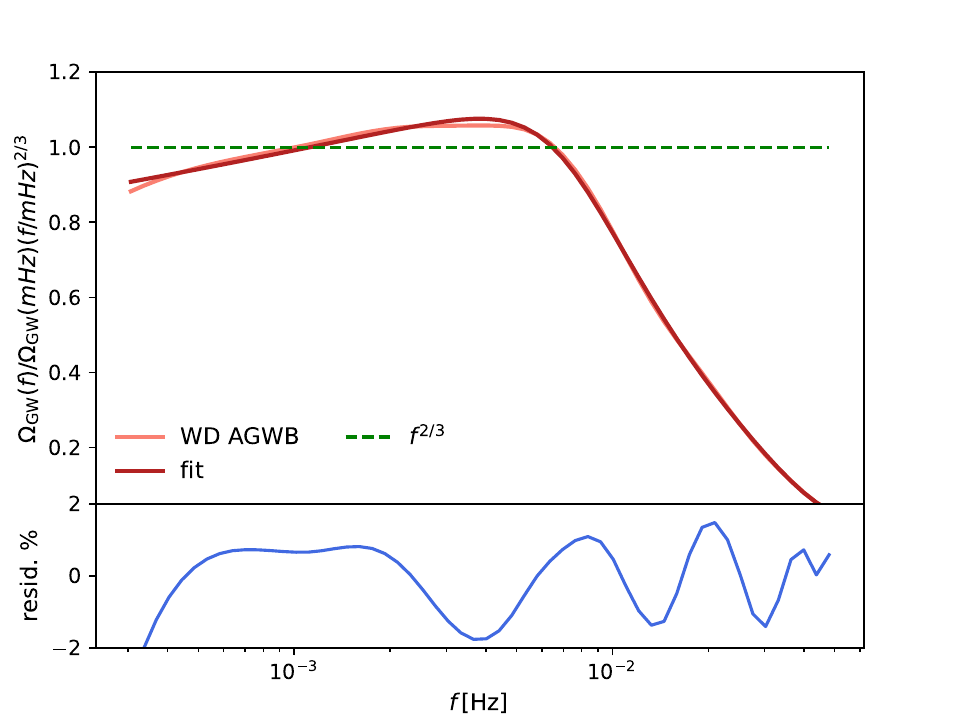}
    \caption{Comparison of the WD AGWB (salmon) and the broken power law fit (dark red) to a purely $f^{2/3}$ signal (green dashed) by dividing the curves by $f^{2/3}$. The WD AGWB deviates from the expected  $2/3$  slope but the residuals (bottom) between the WD AGWB and the fit are below 2\%.}
   \label{fig:fit}
\end{figure}

We show the final results of this work in Fig.~\ref{fig:Omega final result}, where the standard components are as in Fig.~\ref{fig:Omega ga4 SFH madau&dickinson}. The red line shows the results from the previous work as shown in \citetalias{2024A&A...683A.139S}. In that work, a population model of $\alpha\alpha$ with $\alpha$ = 4 and a metallicity of 0.02, as well as the SFRD by \citet{madau_cosmic_2014} were used. The light green corresponds to the WD component of the AGWB for our standard model $\gamma\alpha$ with $\alpha$ = 4 and the MZ19 SFRD model. The light green band around the AGWB signal represents the uncertainties that arose in this work. We find these uncertainties by determining the upper and lower limits of the AGWB signal at all frequencies that come from the use of the different metallicity dependent SFRD models combined with all four binary population models. We find a strength of the WD component of the AGWB at 1 mHz of $\Omega_{\mathrm{WD}}(1 \mathrm{mHz}) = 4.01_{-2.03}^{+5.63} \times 10^{-12}$. Just before the turnover, at a frequency of about 7 mHz, we find the following results: $\Omega_{\mathrm{WD}}(7 \mathrm{mHz}) = 1.43_{-0.68}^{+1.95} \times 10^{-11}$. Looking at the factor between the upper and lower limits of the AGWB signal at 1 mHz and 7 mHz, respectively, we find a factor of approximately 5 difference between the upper and the lower limits. We also show the normal LISA sensitivity, which is above the WD AGWB, to make clear that the WD AGWB does not significantly impact the other LISA science.



We find that the AGWB is comprised of approximately $2.44\times 10^{17}$ binaries, which is roughly in agreement with \citetalias{2024A&A...683A.139S} (who find approximately $1.5\times 10^{17}$ binaries). Moreover, we also find a similar redshift contribution as the one found in \citetalias{2024A&A...683A.139S}, with most of the signal originating at redshifts smaller than 1.

As in \citetalias{2024A&A...683A.139S} we make a fit to the resulting WD AGWB. The new standard result has very similar shape so we can use the same broken broken power law with an exponential cutoff:
\begin{align}
    \Omega_{\rm GW}(f) = & A\, \left(\frac{f}{\hat{f}}\right)^{0.741}  \left[1 + \left(\frac{f}{\hat{f}}\right)^{4.15}\right]^{-0.255} \cdot \exp\left(-B f^3\right) \label{eq: fit}
\end{align}
with different powers in the equation and different parameters $ A= 1.72\times 10^{-11}, B =  1.54\times 10^4$ and $\hat{f} = 7.2$ mHz.
In Fig.~\ref{fig:fit} we plot the signal in the region above 0.3 mHz divided by a pure $f^{2/3}$ power law normalised around 1 mHz, showing a similar deviation from the expected $f^{2/3}$ signal as in \citetalias{2024A&A...683A.139S}. The bottom panel shows the residuals of the signal with respect to the fit, which remain below 2 percent.

\section{Discussion and conclusions}\label{discussion}

In this study we explored some of the factors that introduce an uncertainty in the expected level and shape of the WD AGWB. Although this gives a good indication of the range of possibilities that are definitely possible, it is not an exhaustive study of all the possible effects. For the SFRD models the extensive survey of the existing observational evidence done by \citet{2019MNRAS.488.5300C,2020A&A...636A..10C,2021MNRAS.508.4994C} implies that the largest uncertainties are in fact captured by the different models we use. However, for the binary population models this is less clear. The large number of uncertain phases in the evolution of binaries makes a quantitative assessment of the uncertainties difficult. We also still neglected the possibility that metallicity also changes the initial binary parameters 
\citep{2018ApJ...854..147B,2019ApJ...875...61M}. Comparison of the same binary models with the very local population of double WD in the Solar neighbourhood \citep[e.g][]{2001A&A...365..491N,2012A&A...546A..70T} suggests that at least at approximately Solar metallicity the models are not too far off, albeit if this means a factor 3 or a factor 5 is unclear.

We compare our main results to the results in \citet{2003MNRAS.346.1197F} and \citetalias{2024A&A...683A.139S}. They find values for $\Omega_{\mathrm{WD}}$ at a frequency of 1 mHz of: $\Omega_{\mathrm{WD}}(1 \mathrm{mHz}) = 3.57\times 10^{-12}$ and $\Omega_{\mathrm{WD}}(1 \mathrm{mHz}) = 5.7\times 10^{-12}$, respectively. We find $\Omega_{\mathrm{WD}}(1 \mathrm{mHz}) = 4.01\times 10^{-12}$. This difference is not unexpected, since the WD component of the AGWB as shown in Fig.~\ref{fig:Omega final result} shows the result of \citetalias{2024A&A...683A.139S} above the result of this work. Possible explanations for the differences between \citet{2003MNRAS.346.1197F} and \citetalias{2024A&A...683A.139S} are described in the latter. Differences between this work and \citetalias{2024A&A...683A.139S} can be explained by both the usage of a different population model as well as a different SFRD model.

There are some simplifications that we make in the calculation, in particular about the population of sources that give rise to the AGWB being completely isotropic. At large redshifts this is very likely completely justified, but as a significant contribution to the signal originates within the volume smaller than z = 0.5, the total signal will be anisotropic since the mass distribution in the Universe is highly structured. It would be interesting to calculate the level of anisotropy and the ability of the LISA measurements to detect this \citep[see][]{2018PhRvL.120w1101C}.

The other question is if LISA can detect the turnover of the WD AGWB around 10 mHz, which seems likely given the sensitivity curve, but also how well the shape of the turnover can be characterised. Even more interesting would of course be the detection of the second turnover around 50 mHz where the BH AGWB is expected to become dominant over the WD AGWB.

In conclusion, we investigated the influence of metallicity, different star formation estimates across cosmic time and different descriptions of the common-envelope phase on the level and shape of the WD AGWB. We conclude that the metallicity does not significantly impact the results, since only the very lowest metallicity (Z = 0.0001) produces a lower signal. \g{However, the uncertainties in the (total amount) of star formation give rise to an significant uncertainty in the level of the signal, while the different binary populations have a smaller effect on the level, but also influence the shape of the turnover of the signal. The total uncertainty on the level of the signal is about a factor of 5.} The uncertainty in the model suggests detection of the signal by a future GW detector would provide information about the star formation history and the binary properties at significant redshifts. The uncertainties are not large enough to challenge the conclusion that in the mHz regime the AGWB will be dominated by the WD population, significantly above the contribution of BHs.


\begin{acknowledgements}
We thank Seppe Staelens for his earlier work, discussions and help \g{and the referee for valuable suggestions and comments}. G.N. is supported by the Dutch science foundation NWO.
\end{acknowledgements}

\bibliographystyle{aa} 
\bibliography{WD_AGWB} 

\begin{thebibliography}{58}
\expandafter\ifx\csname natexlab\endcsname\relax\def\natexlab#1{#1}\fi

\bibitem[{{Abbott} {et~al.}(2021){Abbott}, {Abbott}, {Abraham}, {Acernese},
  {Ackley}, {Adams}, {Adams}, {Adhikari}, {Adya}, {Affeldt}, {Agarwal},
  {Agathos}, {Agatsuma}, {Aggarwal}, {Aguiar}, {Aiello}, {Ain}, {Akutsu},
  {Aleman}, {Allen}, {Allocca}, {Altin}, {Amato}, {Anand}, {Ananyeva},
  {Anderson}, {Anderson}, {Ando}, {Angelova}, {Ansoldi}, {Antelis}, {Antier},
  {Appert}, {Arai}, {Arai}, {Arai}, {Araki}, {Araya}, {Araya}, {Areeda},
  {Ar{\`e}ne}, {Aritomi}, {Arnaud}, {Aronson}, {Asada}, {Asali}, {Ashton},
  {Aso}, {Aston}, {Astone}, {Aubin}, {Aufmuth}, {Aultoneal}, {Austin}, {Babak},
  {Badaracco}, {Bader}, {Bae}, {Bae}, {Baer}, {Bagnasco}, {Bai}, {Baiotti},
  {Baird}, {Bajpai}, {Ball}, {Ballardin}, {Ballmer}, {Bals}, {Balsamo},
  {Baltus}, {Banagiri}, {Bankar}, {Bankar}, {Barayoga}, {Barbieri}, {Barish},
  {Barker}, {Barneo}, {Barnum}, {Barone}, {Barr}, {Barsotti}, {Barsuglia},
  {Barta}, {Bartlett}, {Barton}, {Bartos}, {Bassiri}, {Basti}, {Bawaj},
  {Bayley}, {Baylor}, {Bazzan}, {B{\'e}csy}, {Bedakihale}, {Bejger},
  {Belahcene}, {Benedetto}, {Beniwal}, {Benjamin}, {Bennett}, {Bentley},
  {Benyaala}, {Bergamin}, {Berger}, {Bernuzzi}, {Bersanetti}, {Bertolini},
  {Betzwieser}, {Bhandare}, {Bhandari}, {Bhattacharjee}, {Bhaumik}, {Bidler},
  {Bilenko}, {Billingsley}, {Birney}, {Birnholtz}, {Biscans}, {Bischi},
  {Biscoveanu}, {Bisht}, {Biswas}, {Bitossi}, {Bizouard}, {Blackburn},
  {Blackman}, {Blair}, {Blair}, {Blair}, {Bobba}, {Bode}, {Boer}, {Bogaert},
  {Boldrini}, {Bondu}, {Bonilla}, {Bonnand}, {Booker}, {Boom}, {Bork},
  {Boschi}, {Bose}, {Bose}, {Bossilkov}, {Boudart}, {Bouffanais}, {Bozzi},
  {Bradaschia}, {Brady}, {Bramley}, {Branch}, {Branchesi}, {Brau}, {Breschi},
  {Briant}, {Briggs}, {Brillet}, {Brinkmann}, {Brockill}, {Brooks}, {Brooks},
  {Brown}, {Brunett}, {Bruno}, {Bruntz}, {Bryant}, {Buikema}, {Bulik},
  {Bulten}, {Buonanno}, {Buscicchio}, {Buskulic}, {Byer}, {Cadonati}, {Caesar},
  {Cagnoli}, {Cahillane}, {Cain}, {Bustillo}, {Callaghan}, {Callister},
  {Calloni}, {Camp}, {Canepa}, {Cannavacciuolo}, {Cannon}, {Cao}, {Cao}, {Cao},
  {Capocasa}, {Capote}, {Carapella}, {Carbognani}, {Carlin}, {Carney},
  {Carpinelli}, {Carullo}, {Carver}, {Diaz}, {Casentini}, {Castaldi},
  {Caudill}, {Cavagli{\`a}}, {Cavalier}, {Cavalieri}, {Cella},
  {Cerd{\'a}-Dur{\'a}n}, {Cesarini}, {Chaibi}, {Chakravarti}, {Champion},
  {Chan}, {Chan}, {Chan}, {Chan}, {Chandra}, {Chanial}, {Chao}, {Charlton},
  {Chase}, {Chassande-Mottin}, {Chatterjee}, {Chaturvedi}, {Chen}, {Chen},
  {Chen}, {Chen}, {Chen}, {Chen}, {Chen}, {Chen}, {Chen}, {Cheng}, {Cheong},
  {Cheung}, {Chia}, {Chiadini}, {Chiang}, {Chierici}, {Chincarini}, {Chiofalo},
  {Chiummo}, {Cho}, {Cho}, {Choate}, {Choudhary}, {Choudhary}, {Christensen},
  {Chu}, {Chu}, {Chu}, {Chua}, {Chung}, {Ciani}, {Ciecielag}, {Cie{\'s}lar},
  {Cifaldi}, {Ciobanu}, {Ciolfi}, {Cipriano}, {Cirone}, {Clara}, {Clark},
  {Clark}, {Clarke}, {Clearwater}, {Clesse}, {Cleva}, {Coccia}, {Cohadon},
  {Cohen}, {Cohen}, {Colleoni}, {Collette}, {Colpi}, {Compton}, {Constancio},
  {Conti}, {Cooper}, {Corban}, {Corbitt}, {Cordero-Carri{\'o}n}, {Corezzi},
  {Corley}, {Cornish}, {Corre}, {Corsi}, {Cortese}, {Costa}, {Cotesta},
  {Coughlin}, {Coughlin}, {Coulon}, {Countryman}, {Cousins}, {Couvares},
  {Covas}, {Coward}, {Cowart}, {Coyne}, {Coyne}, {Creighton}, {Creighton},
  {Criswell}, {Croquette}, {Crowder}, {Cudell}, {Cullen}, {Cumming},
  {Cummings}, {Cuoco}, {Cury{\l}o}, {Canton}, {D{\'a}lya}, {Dana},
  {Daneshgaranbajastani}, {D'Angelo}, {Danilishin}, {D'Antonio}, {Danzmann},
  {Darsow-Fromm}, {Dasgupta}, {Datrier}, {Dattilo}, {Dave}, {Davier}, {Davies},
  {Davis}, {Daw}, {Dean}, {Deenadayalan}, {Degallaix}, {de Laurentis},
  {Del{\'e}glise}, {Del Favero}, {de Lillo}, {de Lillo}, {Del Pozzo},
  {Demarchi}, {de Matteis}, {D'Emilio}, {Demos}, {Dent}, {Depasse}, {de
  Pietri}, {De Rosa}, {de Rossi}, {Desalvo}, {de Simone}, {Dhurandhar},
  {D{\'\i}az}, {Diaz-Ortiz}, {Didio}, {Dietrich}, {di Fiore}, {di Fronzo}, {di
  Giorgio}, {di Giovanni}, {di Girolamo}, {di Lieto}, {Ding}, {di Pace}, {di
  Palma}, {di Renzo}, {Divakarla}, {Dmitriev}, {Doctor}, {D'Onofrio},
  {Donovan}, {Dooley}, {Doravari}, {Dorrington}, {Drago}, {Driggers}, {Drori},
  {Du}, {Ducoin}, {Dupej}, {Durante}, {D'Urso}, {Duverne}, {Dvorkin}, {Dwyer},
  {Easter}, {Ebersold}, {Eddolls}, {Edelman}, {Edo}, {Edy}, {Effler}, {Eguchi},
  {Eichholz}, {Eikenberry}, {Eisenmann}, {Eisenstein}, {Ejlli}, {Enomoto},
  {Errico}, {Essick}, {Estell{\'e}s}, {Estevez}, {Etienne}, {Etzel}, {Evans},
  {Evans}, {Ewing}, {Fafone}, {Fair}, {Fairhurst}, {Fan}, {Farah}, {Farinon},
  {Farr}, {Farr}, {Farrow}, {Fauchon-Jones}, {Favata}, {Fays}, {Fazio},
  {Feicht}, {Fejer}, {Feng}, {Fenyvesi}, {Ferguson}, {Fernandez-Galiana},
  {Ferrante}, {Ferreira}, {Fidecaro}, {Figura}, {Fiori}, {Fishbach}, {Fisher},
  {Fishner}, {Fittipaldi}, {Fiumara}, {Flaminio}, {Floden}, {Flynn}, {Fong},
  {Font}, {Fornal}, {Forsyth}, {Franke}, {Frasca}, {Frasconi}, {Frederick},
  {Frei}, {Freise}, {Frey}, {Fritschel}, {Frolov}, {Fronz{\'e}}, {Fujii},
  {Fujikawa}, {Fukunaga}, {Fukushima}, {Fulda}, {Fyffe}, {Gabbard}, {Gadre},
  {Gaebel}, {Gair}, {Gais}, {Galaudage}, {Gamba}, {Ganapathy}, {Ganguly},
  {Gao}, {Gaonkar}, {Garaventa}, {Garc{\'\i}a-N{\'u}{\~n}ez},
  {Garc{\'\i}a-Quir{\'o}s}, {Garufi}, {Gateley}, {Gaudio}, {Gayathri}, {Ge},
  {Gemme}, {Gennai}, {George}, {Gergely}, {Gewecke}, {Ghonge}, {Ghosh},
  {Ghosh}, {Ghosh}, {Ghosh}, {Ghosh}, {Giacomazzo}, {Giacoppo}, {Giaime},
  {Giardina}, {Gibson}, {Gier}, {Giesler}, {Giri}, {Gissi}, {Glanzer},
  {Gleckl}, {Godwin}, {Goetz}, {Goetz}, {Gohlke}, {Goncharov}, {Gonz{\'a}lez},
  {Gopakumar}, {Gosselin}, {Gouaty}, {Grace}, {Grado}, {Granata}, {Granata},
  {Grant}, {Gras}, {Grassia}, {Gray}, {Gray}, {Greco}, {Green}, {Green},
  {Gretarsson}, {Gretarsson}, {Griffith}, {Griffiths}, {Griggs}, {Grignani},
  {Grimaldi}, {Grimes}, {Grimm}, {Grote}, {Grunewald}, {Gruning}, {Guerrero},
  {Guidi}, {Guimaraes}, {Guix{\'e}}, {Gulati}, {Guo}, {Guo}, {Gupta}, {Gupta},
  {Gupta}, {Gustafson}, {Gustafson}, {Guzman}, {Ha}, {Haegel}, {Hagiwara},
  {Haino}, {Halim}, {Hall}, {Hamilton}, {Hammond}, {Han}, {Haney}, {Hanks},
  {Hanna}, {Hannam}, {Hannuksela}, {Hansen}, {Hansen}, {Hanson}, {Harder},
  {Hardwick}, {Haris}, {Harms}, {Harry}, {Harry}, {Hartwig}, {Hasegawa},
  {Haskell}, {Hasskew}, {Haster}, {Hattori}, {Haughian}, {Hayakawa}, {Hayama},
  {Hayes}, {Healy}, {Heidmann}, {Heintze}, {Heinze}, {Heinzel}, {Heitmann},
  {Hellman}, {Hello}, {Helmling-Cornell}, {Hemming}, {Hendry}, {Heng},
  {Hennes}, {Hennig}, {Hennig}, {Vivanco}, {Heurs}, {Hild}, {Hill}, {Himemoto},
  {Hines}, {Hiranuma}, {Hirata}, {Hirose}, {Hochheim}, {Hofman}, {Hohmann},
  {Holgado}, {Holland}, {Hollows}, {Holmes}, {Holt}, {Holz}, {Hong}, {Hopkins},
  {Hough}, {Howell}, {Hoy}, {Hoyland}, {Hreibi}, {Hsieh}, {Hsu}, {Huang},
  {Huang}, {Huang}, {Huang}, {Huang}, {Huang}, {H{\"u}bner}, {Huddart},
  {Huerta}, {Hughey}, {Hui}, {Hui}, {Husa}, {Huttner}, {Huxford}, {Huynh-Dinh},
  {Ide}, {Idzkowski}, {Iess}, {Ikenoue}, {Imam}, {Inayoshi}, {Inchauspe},
  {Ingram}, {Inoue}, {Intini}, {Ioka}, {Isi}, {Isleif}, {Ito}, {Itoh}, {Iyer},
  {Izumi}, {Jaberianhamedan}, {Jacqmin}, {Jadhav}, {Jadhav}, {James}, {Jan},
  {Jani}, {Janssens}, {Janthalur}, {Jaranowski}, {Jariwala}, {Jaume},
  {Jenkins}, {Jeon}, {Jeunon}, {Jia}, {Jiang}, {Jin}, {Johns}, {Jones},
  {Jones}, {Jones}, {Jones}, {Jones}, {Jonker}, {Ju}, {Jung}, {Jung}, {Junker},
  {Kaihotsu}, {Kajita}, {Kakizaki}, {Kalaghatgi}, {Kalogera}, {Kamai},
  {Kamiizumi}, {Kanda}, {Kandhasamy}, {Kang}, {Kanner}, {Kao}, {Kapadia},
  {Kapasi}, {Karathanasis}, {Karki}, {Kashyap}, {Kasprzack}, {Kastaun},
  {Katsanevas}, {Katsavounidis}, {Katzman}, {Kaur}, {Kawabe}, {Kawaguchi},
  {Kawai}, {Kawasaki}, {K{\'e}f{\'e}lian}, {Keitel}, {Key}, {Khadka},
  {Khalili}, {Khan}, {Khan}, {Khazanov}, {Khetan}, {Khursheed}, {Kijbunchoo},
  {Kim}, {Kim}, {Kim}, {Kim}, {Kim}, {Kim}, {Kimball}, {Kimura}, {King},
  {Kinley-Hanlon}, {Kirchhoff}, {Kissel}, {Kita}, {Kitazawa}, {Kleybolte},
  {Klimenko}, {Knee}, {Knowles}, {Knyazev}, {Koch}, {Koekoek}, {Kojima},
  {Kokeyama}, {Koley}, {Kolitsidou}, {Kolstein}, {Komori}, {Kondrashov},
  {Kong}, {Kontos}, {Koper}, {Korobko}, {Kotake}, {Kovalam}, {Kozak},
  {Kozakai}, {Kozu}, {Kringel}, {Krishnendu}, {Kr{\'o}lak}, {Kuehn}, {Kuei},
  {Kumar}, {Kumar}, {Kumar}, {Kumar}, {Kume}, {Kuns}, {Kuo}, {Kuo}, {Kuromiya},
  {Kuroyanagi}, {Kusayanagi}, {Kwak}, {Kwang}, {Laghi}, {Lalande}, {Lam},
  {Lamberts}, {Landry}, {Lane}, {Lang}, {Lange}, {Lantz}, {La Rosa},
  {Lartaux-Vollard}, {Lasky}, {Laxen}, {Lazzarini}, {Lazzaro}, {Leaci},
  {Leavey}, {Lecoeuche}, {Lee}, {Lee}, {Lee}, {Lee}, {Lee}, {Lee}, {Lehmann},
  {Lema{\^\i}tre}, {Leon}, {Leonardi}, {Leroy}, {Letendre}, {Levin}, {Leviton},
  {Li}, {Li}, {Li}, {Li}, {Li}, {Li}, {Lin}, {Lin}, {Lin}, {Lin}, {Lin},
  {Linde}, {Linker}, {Linley}, {Littenberg}, {Liu}, {Liu}, {Liu}, {Liu},
  {Llorens-Monteagudo}, {Lo}, {Lockwood}, {Lollie}, {London}, {Longo}, {Lopez},
  {Lorenzini}, {Loriette}, {Lormand}, {Losurdo}, {Lough}, {Lousto}, {Lovelace},
  {L{\"u}ck}, {Lumaca}, {Lundgren}, {Luo}, {Macas}, {Macinnis}, {MacLeod},
  {MacMillan}, {Macquet}, {Hernandez}, {Maga{\~n}a-Sandoval}, {Magazz{\`u}},
  {Magee}, {Maggiore}, {Majorana}, {Maksimovic}, {Maliakal}, {Malik}, {Man},
  {Mandic}, {Mangano}, {Mango}, {Mansell}, {Manske}, {Mantovani}, {Mapelli},
  {Marchesoni}, {Marchio}, {Marion}, {Mark}, {M{\'a}rka}, {M{\'a}rka},
  {Markakis}, {Markosyan}, {Markowitz}, {Maros}, {Marquina}, {Marsat},
  {Martelli}, {Martin}, {Martin}, {Martinez}, {Martinez}, {Martinovic},
  {Martynov}, {Marx}, {Masalehdan}, {Mason}, {Massera}, {Masserot},
  {Massinger}, {Masso-Reid}, {Mastrogiovanni}, {Matas}, {Mateu-Lucena},
  {Matichard}, {Matiushechkina}, {Mavalvala}, {McCann}, {McCarthy},
  {McClelland}, {McClincy}, {McCormick}, {McCuller}, {McGhee}, {McGuire},
  {McIsaac}, {McIver}, {McManus}, {McRae}, {McWilliams}, {Meacher}, {Mehmet},
  {Mehta}, {Melatos}, {Melchor}, {Mendell}, {Menendez-Vazquez}, {Menoni},
  {Mercer}, {Mereni}, {Merfeld}, {Merilh}, {Merritt}, {Merzougui}, {Meshkov},
  {Messenger}, {Messick}, {Meyers}, {Meylahn}, {Mhaske}, {Miani}, {Miao},
  {Michaloliakos}, {Michel}, {Michimura}, {Middleton}, {Milano}, {Miller},
  {Millhouse}, {Mills}, {Milotti}, {Milovich-Goff}, {Minazzoli}, {Minenkov},
  {Mio}, {Mir}, {Mishkin}, {Mishra}, {Mishra}, {Mistry}, {Mitra}, {Mitrofanov},
  {Mitselmakher}, {Mittleman}, {Miyakawa}, {Miyamoto}, {Miyazaki}, {Miyo},
  {Miyoki}, {Mo}, {Mogushi}, {Mohapatra}, {Mohite}, {Molina}, {Molina-Ruiz},
  {Mondin}, {Montani}, {Moore}, {Moraru}, {Morawski}, {More}, {Moreno},
  {Moreno}, {Mori}, {Morisaki}, {Moriwaki}, {Mours}, {Mow-Lowry}, {Mozzon},
  {Muciaccia}, {Mukherjee}, {Mukherjee}, {Mukherjee}, {Mukherjee}, {Mukund},
  {Mullavey}, {Munch}, {Mu{\~n}iz}, {Murray}, {Musenich}, {Nadji}, {Nagano},
  {Nagano}, {Nagar}, {Nakamura}, {Nakano}, {Nakano}, {Nakashima}, {Nakayama},
  {Nardecchia}, {Narikawa}, {Naticchioni}, {Nayak}, {Nayak}, {Negishi}, {Neil},
  {Neilson}, {Nelemans}, {Nelson}, {Nery}, {Neunzert}, {Ng}, {Ng}, {Nguyen},
  {Nguyen}, {Nguyen}, {Quynh}, {Ni}, {Nichols}, {Nishizawa}, {Nissanke},
  {Nocera}, {Noh}, {Norman}, {North}, {Nozaki}, {Nuttall}, {Oberling},
  {O'Brien}, {Obuchi}, {O'Dell}, {Ogaki}, {Oganesyan}, {Oh}, {Oh}, {Oh},
  {Ohashi}, {Ohishi}, {Ohkawa}, {Ohme}, {Ohta}, {Okada}, {Okutani}, {Okutomi},
  {Olivetto}, {Oohara}, {Ooi}, {Oram}, {O'Reilly}, {Ormiston}, {Ormsby},
  {Ortega}, {O'Shaughnessy}, {O'Shea}, {Oshino}, {Ossokine}, {Osthelder},
  {Otabe}, {Ottaway}, {Overmier}, {Pace}, {Pagano}, {Page}, {Pagliaroli},
  {Pai}, {Pai}, {Palamos}, {Palashov}, {Palomba}, {Pan}, {Panda}, {Pang},
  {Pang}, {Pankow}, {Pannarale}, {Pant}, {Paoletti}, {Paoli}, {Paolone},
  {Parisi}, {Park}, {Parker}, {Pascucci}, {Pasqualetti}, {Passaquieti},
  {Passuello}, {Patel}, {Patricelli}, {Payne}, {Pechsiri}, {Pedraza},
  {Pegoraro}, {Pele}, {Arellano}, {Penn}, {Perego}, {Pereira}, {Pereira},
  {Perez}, {P{\'e}rigois}, {Perreca}, {Perri{\`e}s}, {Petermann}, {Petterson},
  {Pfeiffer}, {Pham}, {Phukon}, {Piccinni}, {Pichot}, {Piendibene},
  {Piergiovanni}, {Pierini}, {Pierro}, {Pillant}, {Pilo}, {Pinard}, {Pinto},
  {Piotrzkowski}, {Piotrzkowski}, {Pirello}, {Pitkin}, {Placidi}, {Plastino},
  {Pluchar}, {Poggiani}, {Polini}, {Pong}, {Ponrathnam}, {Popolizio}, {Porter},
  {Powell}, {Pracchia}, {Pradier}, {Prajapati}, {Prasai}, {Prasanna},
  {Pratten}, {Prestegard}, {Principe}, {Prodi}, {Prokhorov}, {Prosposito},
  {Prudenzi}, {Puecher}, {Punturo}, {Puosi}, {Puppo}, {P{\"u}rrer}, {Qi},
  {Quetschke}, {Quinonez}, {Quitzow-James}, {Raab}, {Raaijmakers}, {Radkins},
  {Radulesco}, {Raffai}, {Rail}, {Raja}, {Rajan}, {Ramirez}, {Ramirez},
  {Ramos-Buades}, {Rana}, {Rapagnani}, {Rapol}, {Ratto}, {Raymond}, {Raza},
  {Razzano}, {Read}, {Rees}, {Regimbau}, {Rei}, {Reid}, {Reitze}, {Relton},
  {Rettegno}, {Ricci}, {Richardson}, {Richardson}, {Richardson}, {Ricker},
  {Riemenschneider}, {Riles}, {Rizzo}, {Robertson}, {Robie}, {Robinet},
  {Rocchi}, {Rocha}, {Rodriguez}, {Rodriguez-Soto}, {Rolland}, {Rollins},
  {Roma}, {Romanelli}, {Romano}, {Romano}, {Romel}, {Romero}, {Romero-Shaw},
  {Romie}, {Rose}, {Rosi{\'n}ska}, {Rosofsky}, {Ross}, {Rowan}, {Rowlinson},
  {Roy}, {Roy}, {Rozza}, {Ruggi}, {Ryan}, {Sachdev}, {Sadecki}, {Sadiq},
  {Sago}, {Saito}, {Saito}, {Sakai}, {Sakai}, {Sakellariadou}, {Sakuno},
  {Salafia}, {Salconi}, {Saleem}, {Salemi}, {Samajdar}, {Sanchez}, {Sanchez},
  {Sanchez}, {Sanchis-Gual}, {Sanders}, {Sanuy}, {Saravanan}, {Sarin},
  {Sassolas}, {Satari}, {Sato}, {Sato}, {Sauter}, {Savage}, {Savant}, {Sawada},
  {Sawant}, {Sawant}, {Sayah}, {Schaetzl}, {Scheel}, {Scheuer},
  {Schindler-Tyka}, {Schmidt}, {Schnabel}, {Schneewind}, {Schofield},
  {Sch{\"o}nbeck}, {Schulte}, {Schutz}, {Schwartz}, {Scott}, {Scott},
  {Seglar-Arroyo}, {Seidel}, {Sekiguchi}, {Sekiguchi}, {Sellers}, {Sengupta},
  {Sennett}, {Sentenac}, {Seo}, {Sequino}, {Sergeev}, {Setyawati}, {Shaffer},
  {Shahriar}, {Shams}, {Shao}, {Sharifi}, {Sharma}, {Sharma}, {Shawhan},
  {Shcheblanov}, {Shen}, {Shibagaki}, {Shikauchi}, {Shimizu}, {Shimoda},
  {Shimode}, {Shink}, {Shinkai}, {Shishido}, {Shoda}, {Shoemaker}, {Shoemaker},
  {Shukla}, {Shyamsundar}, {Sieniawska}, {Sigg}, {Singer}, {Singh}, {Singh},
  {Singha}, {Sintes}, {Sipala}, {Skliris}, {Slagmolen}, {Slaven-Blair},
  {Smetana}, {Smith}, {Smith}, {Somala}, {Somiya}, {Son}, {Soni}, {Soni},
  {Sorazu}, {Sordini}, {Sorrentino}, {Sorrentino}, {Sotani}, {Soulard},
  {Souradeep}, {Sowell}, {Spagnuolo}, {Spencer}, {Spera}, {Srivastava},
  {Srivastava}, {Staats}, {Stachie}, {Steer}, {Steinlechner}, {Steinlechner},
  {Stops}, {Stover}, {Strain}, {Strang}, {Stratta}, {Strunk}, {Sturani},
  {Stuver}, {S{\"u}dbeck}, {Sudhagar}, {Sudhir}, {Sugimoto}, {Suh},
  {Summerscales}, {Sun}, {Sun}, {Sunil}, {Sur}, {Suresh}, {Sutton}, {Suzuki},
  {Suzuki}, {Swinkels}, {Szczepa{\'n}czyk}, {Szewczyk}, {Tacca}, {Tagoshi},
  {Tait}, {Takahashi}, {Takahashi}, {Takamori}, {Takano}, {Takeda}, {Takeda},
  {Talbot}, {Tanaka}, {Tanaka}, {Tanaka}, {Tanaka}, {Tanaka}, {Tanasijczuk},
  {Tanioka}, {Tanner}, {Tao}, {Tapia}, {Martin}, {Martin}, {Tasson}, {Telada},
  {Tenorio}, {Terkowski}, {Test}, {Thirugnanasambandam}, {Thomas}, {Thomas},
  {Thompson}, {Thondapu}, {Thorne}, {Thrane}, {Tiwari}, {Tiwari}, {Tiwari},
  {Toland}, {Tolley}, {Tomaru}, {Tomigami}, {Tomura}, {Tonelli},
  {Torres-Forn{\'e}}, {Torrie}, {E Melo}, {T{\"o}yr{\"a}}, {Trapananti},
  {Travasso}, {Traylor}, {Tringali}, {Tripathee}, {Troiano}, {Trovato},
  {Trozzo}, {Trudeau}, {Tsai}, {Tsai}, {Tsang}, {Tsang}, {Tsao}, {Tse}, {Tso},
  {Tsubono}, {Tsuchida}, {Tsukada}, {Tsuna}, {Tsutsui}, {Tsuzuki}, {Turconi},
  {Tuyenbayev}, {Ubhi}, {Uchikata}, {Uchiyama}, {Udall}, {Ueda}, {Uehara},
  {Ueno}, {Ueshima}, {Ugolini}, {Unnikrishnan}, {Uraguchi}, {Urban}, {Ushiba},
  {Usman}, {Utina}, {Vahlbruch}, {Vajente}, {Vajpeyi}, {Valdes}, {Valentini},
  {Valsan}, {van Bakel}, {van Beuzekom}, {van den Brand}, {van den Broeck},
  {van Remortel}, {Vander-Hyde}, {van der Schaaf}, {van Heijningen}, {van
  Putten}, {Vardaro}, {Vargas}, {Varma}, {Vas{\'u}th}, {Vecchio}, {Vedovato},
  {Veitch}, {Veitch}, {Venkateswara}, {Venneberg}, {Venugopalan}, {Verkindt},
  {Verma}, {Veske}, {Vetrano}, {Vicer{\'e}}, {Viets}, {Villa-Ortega}, {Vinet},
  {Vitale}, {Vo}, {Vocca}, {von Reis}, {von Wrangel}, {Vorvick}, {Vyatchanin},
  {Wade}, {Wade}, {Wagner}, {Walet}, {Walker}, {Wallace}, {Wallace}, {Walsh},
  {Wang}, {Wang}, {Wang}, {Ward}, {Warner}, {Was}, {Washimi}, {Washington},
  {Watchi}, {Weaver}, {Wei}, {Weinert}, {Weinstein}, {Weiss}, {Weller},
  {Wellmann}, {Wen}, {We{\ss}els}, {Westhouse}, {Wette}, {Whelan}, {White},
  {Whiting}, {Whittle}, {Wilken}, {Williams}, {Williams}, {Williamson},
  {Willis}, {Willke}, {Wilson}, {Winkler}, {Wipf}, {Wlodarczyk}, {Woan},
  {Woehler}, {Wofford}, {Wong}, {Wu}, {Wu}, {Wu}, {Wu}, {Wysocki}, {Xiao},
  {Xu}, {Yamada}, {Yamamoto}, {Yamamoto}, {Yamamoto}, {Yamamoto}, {Yamashita},
  {Yamazaki}, {Yang}, {Yang}, {Yang}, {Yang}, {Yang}, {Yap}, {Yeeles},
  {Yelikar}, {Ying}, {Yokogawa}, {Yokoyama}, {Yokozawa}, {Yoon}, {Yoshioka},
  {Yu}, {Yu}, {Yuzurihara}, {Zadro{\.z}ny}, {Zanolin}, {Zeidler}, {Zelenova},
  {Zendri}, {Zevin}, {Zhan}, {Zhang}, {Zhang}, {Zhang}, {Zhang}, {Zhang},
  {Zhao}, {Zhao}, {Zhao}, {Zhao}, {Zhou}, {Zhu}, {Zhu}, {Zucker}, {Zweizig},
  {Ligo Scientific Collaboration}, {VIRGO Collaboration}, \& {Kagra
  Collaboration}}]{2021PhRvD.104b2004A}
{Abbott}, R., {Abbott}, T.~D., {Abraham}, S., {et~al.} 2021, \prd, 104, 022004

\bibitem[{{Abbott} {et~al.}(2023){Abbott}, {Abbott}, {Acernese}, {Ackley},
  {Adams}, {Adhikari}, {Adhikari}, {Adya}, {Affeldt}, {Agarwal}, {Agathos},
  {Agatsuma}, {Aggarwal}, {Aguiar}, {Aiello}, {Ain}, {Ajith}, {Akutsu}, {de
  Alarc{\'o}n}, {Akcay}, {Albanesi}, {Allocca}, {Altin}, {Amato}, {Anand},
  {Anand}, {Ananyeva}, {Anderson}, {Anderson}, {Ando}, {Andrade}, {Andres},
  {Andri{\'c}}, {Angelova}, {Ansoldi}, {Antelis}, {Antier}, {Antonini},
  {Appert}, {Arai}, {Arai}, {Arai}, {Araki}, {Araya}, {Araya}, {Areeda},
  {Ar{\`e}ne}, {Aritomi}, {Arnaud}, {Arogeti}, {Aronson}, {Arun}, {Asada},
  {Asali}, {Ashton}, {Aso}, {Assiduo}, {Aston}, {Astone}, {Aubin}, {Austin},
  {Babak}, {Badaracco}, {Bader}, {Badger}, {Bae}, {Bae}, {Baer}, {Bagnasco},
  {Bai}, {Baiotti}, {Baird}, {Bajpai}, {Ball}, {Ballardin}, {Ballmer},
  {Balsamo}, {Baltus}, {Banagiri}, {Bankar}, {Barayoga}, {Barbieri}, {Barish},
  {Barker}, {Barneo}, {Barone}, {Barr}, {Barsotti}, {Barsuglia}, {Barta},
  {Bartlett}, {Barton}, {Bartos}, {Bassiri}, {Basti}, {Bawaj}, {Bayley},
  {Baylor}, {Bazzan}, {B{\'e}csy}, {Bedakihale}, {Bejger}, {Belahcene},
  {Benedetto}, {Beniwal}, {Bennett}, {Bentley}, {Benyaala}, {Bergamin},
  {Berger}, {Bernuzzi}, {Berry}, {Bersanetti}, {Bertolini}, {Betzwieser},
  {Beveridge}, {Bhandare}, {Bhardwaj}, {Bhattacharjee}, {Bhaumik}, {Bilenko},
  {Billingsley}, {Bini}, {Birney}, {Birnholtz}, {Biscans}, {Bischi},
  {Biscoveanu}, {Bisht}, {Biswas}, {Bitossi}, {Bizouard}, {Blackburn}, {Blair},
  {Blair}, {Blair}, {Bobba}, {Bode}, {Boer}, {Bogaert}, {Boldrini}, {Bonavena},
  {Bondu}, {Bonilla}, {Bonnand}, {Booker}, {Boom}, {Bork}, {Boschi}, {Bose},
  {Bose}, {Bossilkov}, {Boudart}, {Bouffanais}, {Bozzi}, {Bradaschia}, {Brady},
  {Bramley}, {Branch}, {Branchesi}, {Brandt}, {Brau}, {Breschi}, {Briant},
  {Briggs}, {Brillet}, {Brinkmann}, {Brockill}, {Brooks}, {Brooks}, {Brown},
  {Brunett}, {Bruno}, {Bruntz}, {Bryant}, {Bulik}, {Bulten}, {Buonanno},
  {Buscicchio}, {Buskulic}, {Buy}, {Byer}, {Cadonati}, {Cagnoli}, {Cahillane},
  {Bustillo}, {Callaghan}, {Callister}, {Calloni}, {Cameron}, {Camp}, {Canepa},
  {Canevarolo}, {Cannavacciuolo}, {Cannon}, {Cao}, {Cao}, {Capocasa}, {Capote},
  {Carapella}, {Carbognani}, {Carlin}, {Carney}, {Carpinelli}, {Carrillo},
  {Carullo}, {Carver}, {Diaz}, {Casentini}, {Castaldi}, {Caudill},
  {Cavagli{\`a}}, {Cavalier}, {Cavalieri}, {Ceasar}, {Cella},
  {Cerd{\'a}-Dur{\'a}n}, {Cesarini}, {Chaibi}, {Chakravarti}, {Subrahmanya},
  {Champion}, {Chan}, {Chan}, {Chan}, {Chan}, {Chan}, {Chandra}, {Chanial},
  {Chao}, {Chapman-Bird}, {Charlton}, {Chase}, {Chassande-Mottin},
  {Chatterjee}, {Chatterjee}, {Chatterjee}, {Chaturvedi}, {Chaty},
  {Chatziioannou}, {Chen}, {Chen}, {Chen}, {Chen}, {Chen}, {Chen}, {Chen},
  {Chen}, {Cheng}, {Cheong}, {Cheung}, {Chia}, {Chiadini}, {Chiang},
  {Chiarini}, {Chierici}, {Chincarini}, {Chiofalo}, {Chiummo}, {Cho}, {Cho},
  {Choudhary}, {Choudhary}, {Christensen}, {Chu}, {Chu}, {Chu}, {Chua},
  {Chung}, {Ciani}, {Ciecielag}, {Cie{\'s}lar}, {Cifaldi}, {Ciobanu}, {Ciolfi},
  {Cipriano}, {Cirone}, {Clara}, {Clark}, {Clark}, {Clarke}, {Clearwater},
  {Clesse}, {Cleva}, {Coccia}, {Codazzo}, {Cohadon}, {Cohen}, {Cohen},
  {Colleoni}, {Collette}, {Colombo}, {Colpi}, {Compton}, {Constancio}, {Conti},
  {Cooper}, {Corban}, {Corbitt}, {Cordero-Carri{\'o}n}, {Corezzi}, {Corley},
  {Cornish}, {Corre}, {Corsi}, {Cortese}, {Costa}, {Cotesta}, {Coughlin},
  {Coulon}, {Countryman}, {Cousins}, {Couvares}, {Coward}, {Cowart}, {Coyne},
  {Coyne}, {Creighton}, {Creighton}, {Criswell}, {Croquette}, {Crowder},
  {Cudell}, {Cullen}, {Cumming}, {Cummings}, {Cunningham}, {Cuoco},
  {Cury{\l}o}, {Dabadie}, {Canton}, {Dall'Osso}, {D{\'a}lya}, {Dana},
  {Daneshgaranbajastani}, {D'Angelo}, {Danila}, {Danilishin}, {D'Antonio},
  {Danzmann}, {Darsow-Fromm}, {Dasgupta}, {Datrier}, {Datta}, {Dattilo},
  {Dave}, {Davier}, {Davies}, {Davis}, {Davis}, {Daw}, {Dean}, {Debra},
  {Deenadayalan}, {Degallaix}, {de Laurentis}, {Del{\'e}glise}, {Del Favero},
  {de Lillo}, {de Lillo}, {Del Pozzo}, {Demarchi}, {de Matteis}, {D'Emilio},
  {Demos}, {Dent}, {Depasse}, {de Pietri}, {De Rosa}, {de Rossi}, {Desalvo},
  {de Simone}, {Dhurandhar}, {D{\'\i}az}, {Diaz-Ortiz}, {Didio}, {Dietrich},
  {di Fiore}, {di Fronzo}, {di Giorgio}, {di Giovanni}, {di Giovanni}, {di
  Girolamo}, {di Lieto}, {Ding}, {di Pace}, {di Palma}, {di Renzo},
  {Divakarla}, {Dmitriev}, {Doctor}, {D'Onofrio}, {Donovan}, {Dooley},
  {Doravari}, {Dorrington}, {Drago}, {Driggers}, {Drori}, {Ducoin}, {Dupej},
  {Durante}, {D'Urso}, {Duverne}, {Dwyer}, {Eassa}, {Easter}, {Ebersold},
  {Eckhardt}, {Eddolls}, {Edelman}, {Edo}, {Edy}, {Effler}, {Eguchi},
  {Eichholz}, {Eikenberry}, {Eisenmann}, {Eisenstein}, {Ejlli}, {Engelby},
  {Enomoto}, {Errico}, {Essick}, {Estell{\'e}s}, {Estevez}, {Etienne}, {Etzel},
  {Evans}, {Evans}, {Ewing}, {Fafone}, {Fair}, {Fairhurst}, {Farah}, {Farinon},
  {Farr}, {Farr}, {Farrow}, {Fauchon-Jones}, {Favaro}, {Favata}, {Fays},
  {Fazio}, {Feicht}, {Fejer}, {Fenyvesi}, {Ferguson}, {Fernandez-Galiana},
  {Ferrante}, {Ferreira}, {Fidecaro}, {Figura}, {Fiori}, {Fishbach}, {Fisher},
  {Fittipaldi}, {Fiumara}, {Flaminio}, {Floden}, {Fong}, {Font}, {Fornal},
  {Forsyth}, {Franke}, {Frasca}, {Frasconi}, {Frederick}, {Freed}, {Frei},
  {Freise}, {Frey}, {Fritschel}, {Frolov}, {Fronz{\'e}}, {Fujii}, {Fujikawa},
  {Fukunaga}, {Fukushima}, {Fulda}, {Fyffe}, {Gabbard}, {Gadre}, {Gair},
  {Gais}, {Galaudage}, {Gamba}, {Ganapathy}, {Ganguly}, {Gao}, {Gaonkar},
  {Garaventa}, {Garc{\'\i}a}, {Garc{\'\i}a-N{\'u}{\~n}ez},
  {Garc{\'\i}a-Quir{\'o}s}, {Garufi}, {Gateley}, {Gaudio}, {Gayathri}, {Ge},
  {Gemme}, {Gennai}, {George}, {George}, {Gerberding}, {Gergely}, {Gewecke},
  {Ghonge}, {Ghosh}, {Ghosh}, {Ghosh}, {Ghosh}, {Giacomazzo}, {Giacoppo},
  {Giaime}, {Giardina}, {Gibson}, {Gier}, {Giesler}, {Giri}, {Gissi},
  {Glanzer}, {Gleckl}, {Godwin}, {Golomb}, {Goetz}, {Goetz}, {Gohlke},
  {Goncharov}, {Gonz{\'a}lez}, {Gopakumar}, {Gosselin}, {Gouaty}, {Gould},
  {Grace}, {Grado}, {Granata}, {Granata}, {Grant}, {Gras}, {Grassia}, {Gray},
  {Gray}, {Greco}, {Green}, {Green}, {Gretarsson}, {Gretarsson}, {Griffith},
  {Griffiths}, {Griggs}, {Grignani}, {Grimaldi}, {Grimm}, {Grote}, {Grunewald},
  {Gruning}, {Guerra}, {Guidi}, {Guimaraes}, {Guix{\'e}}, {Gulati}, {Guo},
  {Guo}, {Gupta}, {Gupta}, {Gupta}, {Gustafson}, {Gustafson}, {Guzman}, {Ha},
  {Haegel}, {Hagiwara}, {Haino}, {Halim}, {Hall}, {Hamilton}, {Hammond}, {Han},
  {Haney}, {Hanks}, {Hanna}, {Hannam}, {Hannuksela}, {Hansen}, {Hansen},
  {Hanson}, {Harder}, {Hardwick}, {Haris}, {Harms}, {Harry}, {Harry},
  {Hartwig}, {Hasegawa}, {Haskell}, {Hasskew}, {Haster}, {Hattori}, {Haughian},
  {Hayakawa}, {Hayama}, {Hayes}, {Healy}, {Heidmann}, {Heidt}, {Heintze},
  {Heinze}, {Heinzel}, {Heitmann}, {Hellman}, {Hello}, {Helmling-Cornell},
  {Hemming}, {Hendry}, {Heng}, {Hennes}, {Hennig}, {Hennig}, {Hernandez},
  {Vivanco}, {Heurs}, {Hild}, {Hill}, {Himemoto}, {Hines}, {Hiranuma},
  {Hirata}, {Hirose}, {Hochheim}, {Hofman}, {Hohmann}, {Holcomb}, {Holland},
  {Hollows}, {Holmes}, {Holt}, {Holz}, {Hong}, {Hopkins}, {Hough}, {Hourihane},
  {Howell}, {Hoy}, {Hoyland}, {Hreibi}, {Hsieh}, {Hsu}, {Huang}, {Huang},
  {Huang}, {Huang}, {Huang}, {Huang}, {H{\"u}bner}, {Huddart}, {Hughey}, {Hui},
  {Hui}, {Husa}, {Huttner}, {Huxford}, {Huynh-Dinh}, {Ide}, {Idzkowski},
  {Iess}, {Ikenoue}, {Imam}, {Inayoshi}, {Ingram}, {Inoue}, {Ioka}, {Isi},
  {Isleif}, {Ito}, {Itoh}, {Iyer}, {Izumi}, {Jaberianhamedan}, {Jacqmin},
  {Jadhav}, {Jadhav}, {James}, {Jan}, {Jani}, {Janquart}, {Janssens},
  {Janthalur}, {Jaranowski}, {Jariwala}, {Jaume}, {Jenkins}, {Jenner}, {Jeon},
  {Jeunon}, {Jia}, {Jin}, {Johns}, {Jones}, {Jones}, {Jones}, {Jones}, {Jones},
  {Jonker}, {Ju}, {Jung}, {Jung}, {Junker}, {Juste}, {Kaihotsu}, {Kajita},
  {Kakizaki}, {Kalaghatgi}, {Kalogera}, {Kamai}, {Kamiizumi}, {Kanda},
  {Kandhasamy}, {Kang}, {Kanner}, {Kao}, {Kapadia}, {Kapasi}, {Karat},
  {Karathanasis}, {Karki}, {Kashyap}, {Kasprzack}, {Kastaun}, {Katsanevas},
  {Katsavounidis}, {Katzman}, {Kaur}, {Kawabe}, {Kawaguchi}, {Kawai},
  {Kawasaki}, {K{\'e}f{\'e}lian}, {Keitel}, {Key}, {Khadka}, {Khalili}, {Khan},
  {Khazanov}, {Khetan}, {Khursheed}, {Kijbunchoo}, {Kim}, {Kim}, {Kim}, {Kim},
  {Kim}, {Kim}, {Kimball}, {Kimura}, {Kinley-Hanlon}, {Kirchhoff}, {Kissel},
  {Kita}, {Kitazawa}, {Kleybolte}, {Klimenko}, {Knee}, {Knowles}, {Knyazev},
  {Koch}, {Koekoek}, {Kojima}, {Kokeyama}, {Koley}, {Kolitsidou}, {Kolstein},
  {Komori}, {Kondrashov}, {Kong}, {Kontos}, {Koper}, {Korobko}, {Kotake},
  {Kovalam}, {Kozak}, {Kozakai}, {Kozu}, {Kringel}, {Krishnendu}, {Kr{\'o}lak},
  {Kuehn}, {Kuei}, {Kuijer}, {Kulkarni}, {Kumar}, {Kumar}, {Kumar}, {Kumar},
  {Kume}, {Kuns}, {Kuo}, {Kuo}, {Kuromiya}, {Kuroyanagi}, {Kusayanagi},
  {Kuwahara}, {Kwak}, {Lagabbe}, {Laghi}, {Lalande}, {Lam}, {Lamberts},
  {Landry}, {Landry}, {Lane}, {Lang}, {Lange}, {Lantz}, {La Rosa},
  {Lartaux-Vollard}, {Lasky}, {Laxen}, {Lazzarini}, {Lazzaro}, {Leaci},
  {Leavey}, {Lecoeuche}, {Lee}, {Lee}, {Lee}, {Lee}, {Lee}, {Lee}, {Lehmann},
  {Lema{\^\i}tre}, {Leonardi}, {Leroy}, {Letendre}, {Levesque}, {Levin},
  {Leviton}, {Leyde}, {Li}, {Li}, {Li}, {Li}, {Li}, {Li}, {Lin}, {Lin}, {Lin},
  {Lin}, {Lin}, {Linde}, {Linker}, {Linley}, {Littenberg}, {Liu}, {Liu}, {Liu},
  {Liu}, {Llamas}, {Llorens-Monteagudo}, {Lo}, {Lockwood}, {Loh}, {London},
  {Longo}, {Lopez}, {Portilla}, {Lorenzini}, {Loriette}, {Lormand}, {Losurdo},
  {Lott}, {Lough}, {Lousto}, {Lovelace}, {Lucaccioni}, {L{\"u}ck}, {Lumaca},
  {Lundgren}, {Luo}, {Lynam}, {Macas}, {Macinnis}, {MacLeod}, {MacMillan},
  {Macquet}, {Hernandez}, {Magazz{\`u}}, {Magee}, {Maggiore}, {Magnozzi},
  {Mahesh}, {Majorana}, {Makarem}, {Maksimovic}, {Maliakal}, {Malik}, {Man},
  {Mandic}, {Mangano}, {Mango}, {Mansell}, {Manske}, {Mantovani}, {Mapelli},
  {Marchesoni}, {Marchio}, {Marion}, {Mark}, {M{\'a}rka}, {M{\'a}rka},
  {Markakis}, {Markosyan}, {Markowitz}, {Maros}, {Marquina}, {Marsat},
  {Martelli}, {Martin}, {Martin}, {Martinez}, {Martinez}, {Martinez},
  {Martinovic}, {Martynov}, {Marx}, {Masalehdan}, {Mason}, {Massera},
  {Masserot}, {Massinger}, {Masso-Reid}, {Mastrogiovanni}, {Matas},
  {Mateu-Lucena}, {Matichard}, {Matiushechkina}, {Mavalvala}, {McCann},
  {McCarthy}, {McClelland}, {McClincy}, {McCormick}, {McCuller}, {McGhee},
  {McGuire}, {McIsaac}, {McIver}, {McRae}, {McWilliams}, {Meacher}, {Mehmet},
  {Mehta}, {Meijer}, {Melatos}, {Melchor}, {Mendell}, {Menendez-Vazquez},
  {Menoni}, {Mercer}, {Mereni}, {Merfeld}, {Merilh}, {Merritt}, {Merzougui},
  {Meshkov}, {Messenger}, {Messick}, {Meyers}, {Meylahn}, {Mhaske}, {Miani},
  {Miao}, {Michaloliakos}, {Michel}, {Michimura}, {Middleton}, {Milano},
  {Miller}, {Miller}, {Miller}, {Miller}, {Millhouse}, {Mills}, {Milotti},
  {Minazzoli}, {Minenkov}, {Mio}, {Mir}, {Miravet-Ten{\'e}s}, {Mishra},
  {Mishra}, {Mistry}, {Mitra}, {Mitrofanov}, {Mitselmakher}, {Mittleman},
  {Miyakawa}, {Miyamoto}, {Miyazaki}, {Miyo}, {Miyoki}, {Mo}, {Modafferi},
  {Moguel}, {Mogushi}, {Mohapatra}, {Mohite}, {Molina}, {Molina-Ruiz},
  {Mondin}, {Montani}, {Moore}, {Moraru}, {Morawski}, {More}, {Moreno},
  {Moreno}, {Mori}, {Morisaki}, {Moriwaki}, {Morr{\'a}s}, {Mours}, {Mow-Lowry},
  {Mozzon}, {Muciaccia}, {Mukherjee}, {Mukherjee}, {Mukherjee}, {Mukherjee},
  {Mukherjee}, {Mukund}, {Mullavey}, {Munch}, {Mu{\~n}iz}, {Murray},
  {Musenich}, {Muusse}, {Nadji}, {Nagano}, {Nagano}, {Nagar}, {Nakamura},
  {Nakano}, {Nakano}, {Nakashima}, {Nakayama}, {Napolano}, {Nardecchia},
  {Narikawa}, {Naticchioni}, {Nayak}, {Nayak}, {Negishi}, {Neil}, {Neilson},
  {Nelemans}, {Nelson}, {Nery}, {Neubauer}, {Neunzert}, {Ng}, {Ng}, {Nguyen},
  {Nguyen}, {Nguyen}, {Quynh}, {Ni}, {Nichols}, {Nishizawa}, {Nissanke},
  {Nitoglia}, {Nocera}, {Norman}, {North}, {Nozaki}, {Siles}, {Nuttall},
  {Oberling}, {O'Brien}, {Obuchi}, {O'Dell}, {Oelker}, {Ogaki}, {Oganesyan},
  {Oh}, {Oh}, {Oh}, {Ohashi}, {Ohishi}, {Ohkawa}, {Ohme}, {Ohta}, {Okada},
  {Okutani}, {Okutomi}, {Olivetto}, {Oohara}, {Ooi}, {Oram}, {O'Reilly},
  {Ormiston}, {Ormsby}, {Ortega}, {O'Shaughnessy}, {O'Shea}, {Oshino},
  {Ossokine}, {Osthelder}, {Otabe}, {Ottaway}, {Overmier}, {Pace}, {Pagano},
  {Page}, {Pagliaroli}, {Pai}, {Pai}, {Palamos}, {Palashov}, {Palomba}, {Pan},
  {Pan}, {Panda}, {Pang}, {Pang}, {Pankow}, {Pannarale}, {Pant}, {Panther},
  {Paoletti}, {Paoli}, {Paolone}, {Parisi}, {Park}, {Park}, {Parker},
  {Pascucci}, {Pasqualetti}, {Passaquieti}, {Passuello}, {Patel}, {Pathak},
  {Patricelli}, {Patron}, {Paul}, {Payne}, {Pedraza}, {Pegoraro}, {Pele},
  {Arellano}, {Penn}, {Perego}, {Pereira}, {Pereira}, {Perez}, {P{\'e}rigois},
  {Perkins}, {Perreca}, {Perri{\`e}s}, {Petermann}, {Petterson}, {Pfeiffer},
  {Pham}, {Phukon}, {Piccinni}, {Pichot}, {Piendibene}, {Piergiovanni},
  {Pierini}, {Pierro}, {Pillant}, {Pillas}, {Pilo}, {Pinard}, {Pinto}, {Pinto},
  {Piotrzkowski}, {Piotrzkowski}, {Pirello}, {Pitkin}, {Placidi}, {Planas},
  {Plastino}, {Pluchar}, {Poggiani}, {Polini}, {Pong}, {Ponrathnam},
  {Popolizio}, {Porter}, {Poulton}, {Powell}, {Pracchia}, {Pradier},
  {Prajapati}, {Prasai}, {Prasanna}, {Pratten}, {Principe}, {Prodi},
  {Prokhorov}, {Prosposito}, {Prudenzi}, {Puecher}, {Punturo}, {Puosi},
  {Puppo}, {P{\"u}rrer}, {Qi}, {Quetschke}, {Quitzow-James}, {Raab},
  {Raaijmakers}, {Radkins}, {Radulesco}, {Raffai}, {Rail}, {Raja}, {Rajan},
  {Ramirez}, {Ramirez}, {Ramos-Buades}, {Rana}, {Rapagnani}, {Rapol}, {Ray},
  {Raymond}, {Raza}, {Razzano}, {Read}, {Rees}, {Regimbau}, {Rei}, {Reid},
  {Reid}, {Reitze}, {Relton}, {Renzini}, {Rettegno}, {Reza}, {Rezac}, {Ricci},
  {Richards}, {Richardson}, {Richardson}, {Riemenschneider}, {Riles},
  {Rinaldi}, {Rink}, {Rizzo}, {Robertson}, {Robie}, {Robinet}, {Rocchi},
  {Rodriguez}, {Rolland}, {Rollins}, {Romanelli}, {Romano}, {Romel},
  {Romero-Rodr{\'\i}guez}, {Romero-Shaw}, {Romie}, {Ronchini}, {Rosa}, {Rose},
  {Rosi{\'n}ska}, {Ross}, {Rowan}, {Rowlinson}, {Roy}, {Roy}, {Roy}, {Rozza},
  {Ruggi}, {Ryan}, {Sachdev}, {Sadecki}, {Sadiq}, {Sago}, {Saito}, {Saito},
  {Sakai}, {Sakai}, {Sakellariadou}, {Sakuno}, {Salafia}, {Salconi}, {Saleem},
  {Salemi}, {Samajdar}, {Sanchez}, {Sanchez}, {Sanchez}, {Sanchis-Gual},
  {Sanders}, {Sanuy}, {Saravanan}, {Sarin}, {Sassolas}, {Satari},
  {Sathyaprakash}, {Sato}, {Sato}, {Sauter}, {Savage}, {Sawada}, {Sawant},
  {Sawant}, {Sayah}, {Schaetzl}, {Scheel}, {Scheuer}, {Schiworski}, {Schmidt},
  {Schmidt}, {Schnabel}, {Schneewind}, {Schofield}, {Sch{\"o}nbeck}, {Schulte},
  {Schutz}, {Schwartz}, {Scott}, {Scott}, {Seglar-Arroyo}, {Sekiguchi},
  {Sekiguchi}, {Sellers}, {Sengupta}, {Sentenac}, {Seo}, {Sequino}, {Sergeev},
  {Setyawati}, {Shaffer}, {Shahriar}, {Shams}, {Shao}, {Sharma}, {Sharma},
  {Shawhan}, {Shcheblanov}, {Shibagaki}, {Shikauchi}, {Shimizu}, {Shimoda},
  {Shimode}, {Shinkai}, {Shishido}, {Shoda}, {Shoemaker}, {Shoemaker},
  {Shyamsundar}, {Sieniawska}, {Sigg}, {Singer}, {Singh}, {Singh}, {Singha},
  {Sintes}, {Sipala}, {Skliris}, {Slagmolen}, {Slaven-Blair}, {Smetana},
  {Smith}, {Smith}, {Soldateschi}, {Somala}, {Somiya}, {Son}, {Soni}, {Soni},
  {Sordini}, {Sorrentino}, {Sorrentino}, {Sotani}, {Soulard}, {Souradeep},
  {Sowell}, {Spagnuolo}, {Spencer}, {Spera}, {Srinivasan}, {Srivastava},
  {Srivastava}, {Staats}, {Stachie}, {Steer}, {Steinhoff}, {Steinlechner},
  {Steinlechner}, {Stevenson}, {Stops}, {Stover}, {Strain}, {Strang},
  {Stratta}, {Strunk}, {Sturani}, {Stuver}, {Sudhagar}, {Sudhir}, {Sugimoto},
  {Suh}, {Sullivan}, {Summerscales}, {Sun}, {Sun}, {Sunil}, {Sur}, {Suresh},
  {Sutton}, {Suzuki}, {Suzuki}, {Swinkels}, {Szczepa{\'n}czyk}, {Szewczyk},
  {Tacca}, {Tagoshi}, {Tait}, {Takahashi}, {Takahashi}, {Takamori}, {Takano},
  {Takeda}, {Takeda}, {Talbot}, {Talbot}, {Tanaka}, {Tanaka}, {Tanaka},
  {Tanaka}, {Tanaka}, {Tanasijczuk}, {Tanioka}, {Tanner}, {Tao}, {Tao},
  {Mart{\'\i}n}, {Taranto}, {Tasson}, {Telada}, {Tenorio}, {Terhune},
  {Terkowski}, {Thirugnanasambandam}, {Thomas}, {Thomas}, {Thomas}, {Thompson},
  {Thondapu}, {Thorne}, {Thrane}, {Tiwari}, {Tiwari}, {Tiwari}, {Toivonen},
  {Toland}, {Tolley}, {Tomaru}, {Tomigami}, {Tomura}, {Tonelli},
  {Torres-Forn{\'e}}, {Torrie}, {E Melo}, {T{\"o}yr{\"a}}, {Trapananti},
  {Travasso}, {Traylor}, {Trevor}, {Tringali}, {Tripathee}, {Troiano},
  {Trovato}, {Trozzo}, {Trudeau}, {Tsai}, {Tsai}, {Tsang}, {Tsang}, {Tsao},
  {Tse}, {Tso}, {Tsubono}, {Tsuchida}, {Tsukada}, {Tsuna}, {Tsutsui},
  {Tsuzuki}, {Turbang}, {Turconi}, {Tuyenbayev}, {Ubhi}, {Uchikata},
  {Uchiyama}, {Udall}, {Ueda}, {Uehara}, {Ueno}, {Ueshima}, {Unnikrishnan},
  {Uraguchi}, {Urban}, {Ushiba}, {Utina}, {Vahlbruch}, {Vajente}, {Vajpeyi},
  {Valdes}, {Valentini}, {Valsan}, {van Bakel}, {van Beuzekom}, {van den
  Brand}, {van den Broeck}, {Vander-Hyde}, {van der Schaaf}, {van Heijningen},
  {Vanosky}, {van Putten}, {van Remortel}, {Vardaro}, {Vargas}, {Varma},
  {Vas{\'u}th}, {Vecchio}, {Vedovato}, {Veitch}, {Veitch}, {Venneberg},
  {Venugopalan}, {Verkindt}, {Verma}, {Verma}, {Veske}, {Vetrano},
  {Vicer{\'e}}, {Vidyant}, {Viets}, {Vijaykumar}, {Villa-Ortega}, {Vinet},
  {Virtuoso}, {Vitale}, {Vo}, {Vocca}, {von Reis}, {von Wrangel}, {Vorvick},
  {Vyatchanin}, {Wade}, {Wade}, {Wagner}, {Walet}, {Walker}, {Wallace},
  {Wallace}, {Walsh}, {Wang}, {Wang}, {Wang}, {Ward}, {Warner}, {Was},
  {Washimi}, {Washington}, {Watchi}, {Weaver}, {Webster}, {Weinert},
  {Weinstein}, {Weiss}, {Weller}, {Wellmann}, {Wen}, {We{\ss}els}, {Wette},
  {Whelan}, {White}, {Whiting}, {Whittle}, {Wilken}, {Williams}, {Williams},
  {Williamson}, {Willis}, {Willke}, {Wilson}, {Winkler}, {Wipf}, {Wlodarczyk},
  {Woan}, {Woehler}, {Wofford}, {Wong}, {Wu}, {Wu}, {Wu}, {Wu}, {Wysocki},
  {Xiao}, {Xu}, {Yamada}, {Yamamoto}, {Yamamoto}, {Yamamoto}, {Yamamoto},
  {Yamashita}, {Yamazaki}, {Yang}, {Yang}, {Yang}, {Yang}, {Yang}, {Yap},
  {Yeeles}, {Yelikar}, {Ying}, {Yokogawa}, {Yokoyama}, {Yokozawa}, {Yoo},
  {Yoshioka}, {Yu}, {Yu}, {Yuzurihara}, {Zadro{\.z}ny}, {Zanolin}, {Zeidler},
  {Zelenova}, {Zendri}, {Zevin}, {Zhan}, {Zhang}, {Zhang}, {Zhang}, {Zhang},
  {Zhang}, {Zhao}, {Zhao}, {Zhao}, {Zhao}, {Zheng}, {Zhou}, {Zhou}, {Zhu},
  {Zhu}, {Zimmerman}, {Zlochower}, {Zucker}, {Zweizig}, {LIGO Scientific
  Collaboration}, {VIRGO Collaboration}, \& {KAGRA
  Collaboration}}]{2023PhRvX..13a1048A}
{Abbott}, R., {Abbott}, T.~D., {Acernese}, F., {et~al.} 2023, Physical Review
  X, 13, 011048

\bibitem[{{Acernese} {et~al.}(2015){Acernese}, {Agathos}, {Agatsuma}, {Aisa},
  {Allemandou}, {Allocca}, {Amarni}, {Astone}, {Balestri}, {Ballardin},
  {Barone}, {Baronick}, {Barsuglia}, {Basti}, {Basti}, {Bauer}, {Bavigadda},
  {Bejger}, {Beker}, {Belczynski}, {Bersanetti}, {Bertolini}, {Bitossi},
  {Bizouard}, {Bloemen}, {Blom}, {Boer}, {Bogaert}, {Bondi}, {Bondu},
  {Bonelli}, {Bonnand}, {Boschi}, {Bosi}, {Bouedo}, {Bradaschia}, {Branchesi},
  {Briant}, {Brillet}, {Brisson}, {Bulik}, {Bulten}, {Buskulic}, {Buy},
  {Cagnoli}, {Calloni}, {Campeggi}, {Canuel}, {Carbognani}, {Cavalier},
  {Cavalieri}, {Cella}, {Cesarini}, {Mottin}, {Chincarini}, {Chiummo}, {Chua},
  {Cleva}, {Coccia}, {Cohadon}, {Colla}, {Colombini}, {Conte}, {Coulon},
  {Cuoco}, {Dalmaz}, {D'Antonio}, {Dattilo}, {Davier}, {Day}, {Debreczeni},
  {Degallaix}, {Del{\'e}glise}, {Pozzo}, {Dereli}, {Rosa}, {Fiore}, {Lieto},
  {Virgilio}, {Doets}, {Dolique}, {Drago}, {Ducrot}, {Endr{\H{o}}czi},
  {Fafone}, {Farinon}, {Ferrante}, {Ferrini}, {Fidecaro}, {Fiori}, {Flaminio},
  {Fournier}, {Franco}, {Frasca}, {Frasconi}, {Gammaitoni}, {Garufi},
  {Gaspard}, {Gatto}, {Gemme}, {Gendre}, {Genin}, {Gennai}, {Ghosh},
  {Giacobone}, {Giazotto}, {Gouaty}, {Granata}, {Greco}, {Groot}, {Guidi},
  {Harms}, {Heidmann}, {Heitmann}, {Hello}, {Hemming}, {Hennes}, {Hofman},
  {Jaranowski}, {Jonker}, {Kasprzack}, {K{\'e}f{\'e}lian}, {Kowalska}, {Kraan},
  {Kr{\'o}lak}, {Kutynia}, {Lazzaro}, {Leonardi}, {Leroy}, {Letendre}, {Li},
  {Lieunard}, {Lorenzini}, {Loriette}, {Losurdo}, {Magazz{\`u}}, {Majorana},
  {Maksimovic}, {Malvezzi}, {Man}, {Mangano}, {Mantovani}, {Marchesoni},
  {Marion}, {Marque}, {Martelli}, {Martellini}, {Masserot}, {Meacher},
  {Meidam}, {Mezzani}, {Michel}, {Milano}, {Minenkov}, {Moggi}, {Mohan},
  {Montani}, {Morgado}, {Mours}, {Mul}, {Nagy}, {Nardecchia}, {Naticchioni},
  {Nelemans}, {Neri}, {Neri}, {Nocera}, {Pacaud}, {Palomba}, {Paoletti},
  {Paoli}, {Pasqualetti}, {Passaquieti}, {Passuello}, {Perciballi}, {Petit},
  {Pichot}, {Piergiovanni}, {Pillant}, {Piluso}, {Pinard}, {Poggiani},
  {Prijatelj}, {Prodi}, {Punturo}, {Puppo}, {Rabeling}, {R{\'a}cz},
  {Rapagnani}, {Razzano}, {Re}, {Regimbau}, {Ricci}, {Robinet}, {Rocchi},
  {Rolland}, {Romano}, {Rosi{\'n}ska}, {Ruggi}, {Saracco}, {Sassolas},
  {Schimmel}, {Sentenac}, {Sequino}, {Shah}, {Siellez}, {Straniero},
  {Swinkels}, {Tacca}, {Tonelli}, {Travasso}, {Turconi}, {Vajente}, {van
  Bakel}, {van Beuzekom}, {van den Brand}, {Van Den Broeck}, {van der Sluys},
  {van Heijningen}, {Vas{\'u}th}, {Vedovato}, {Veitch}, {Verkindt}, {Vetrano},
  {Vicer{\'e}}, {Vinet}, {Visser}, {Vocca}, {Ward}, {Was}, {Wei}, {Yvert},
  {{\.z}ny}, \& {Zendri}}]{2015CQGra..32b4001A}
{Acernese}, F., {Agathos}, M., {Agatsuma}, K., {et~al.} 2015, Classical and
  Quantum Gravity, 32, 024001

\bibitem[{{Alonso} {et~al.}(2020){Alonso}, {Contaldi}, {Cusin}, {Ferreira}, \&
  {Renzini}}]{2020PhRvD.101l4048A}
{Alonso}, D., {Contaldi}, C.~R., {Cusin}, G., {Ferreira}, P.~G., \& {Renzini},
  A.~I. 2020, \prd, 101, 124048

\bibitem[{{Amaro-Seoane} {et~al.}(2023){Amaro-Seoane}, {Andrews}, {Arca Sedda},
  {Askar}, {Baghi}, {Balasov}, {Bartos}, {Bavera}, {Bellovary}, {Berry},
  {Berti}, {Bianchi}, {Blecha}, {Blondin}, {Bogdanovi{\'c}}, {Boissier},
  {Bonetti}, {Bonoli}, {Bortolas}, {Breivik}, {Capelo}, {Caramete},
  {Cattorini}, {Charisi}, {Chaty}, {Chen}, {Chru{\'s}li{\'n}ska}, {Chua},
  {Church}, {Colpi}, {D'Orazio}, {Danielski}, {Davies}, {Dayal}, {De Rosa},
  {Derdzinski}, {Destounis}, {Dotti}, {Dutan}, {Dvorkin}, {Fabj}, {Foglizzo},
  {Ford}, {Fouvry}, {Franchini}, {Fragos}, {Fryer}, {Gaspari}, {Gerosa},
  {Graziani}, {Groot}, {Habouzit}, {Haggard}, {Haiman}, {Han}, {Istrate},
  {Johansson}, {Khan}, {Kimpson}, {Kokkotas}, {Kong}, {Korol}, {Kremer},
  {Kupfer}, {Lamberts}, {Larson}, {Lau}, {Liu}, {Lloyd-Ronning}, {Lodato},
  {Lupi}, {Ma}, {Maccarone}, {Mandel}, {Mangiagli}, {Mapelli}, {Mathis},
  {Mayer}, {McGee}, {McKernan}, {Miller}, {Mota}, {Mumpower}, {Nasim},
  {Nelemans}, {Noble}, {Pacucci}, {Panessa}, {Paschalidis}, {Pfister},
  {Porquet}, {Quenby}, {Ricarte}, {R{\"o}pke}, {Regan}, {Rosswog}, {Ruiter},
  {Ruiz}, {Runnoe}, {Schneider}, {Schnittman}, {Secunda}, {Sesana}, {Seto},
  {Shao}, {Shapiro}, {Sopuerta}, {Stone}, {Suvorov}, {Tamanini}, {Tamfal},
  {Tauris}, {Temmink}, {Tomsick}, {Toonen}, {Torres-Orjuela}, {Toscani},
  {Tsokaros}, {Unal}, {V{\'a}zquez-Aceves}, {Valiante}, {van Putten}, {van
  Roestel}, {Vignali}, {Volonteri}, {Wu}, {Younsi}, {Yu}, {Zane}, {Zwick},
  {Antonini}, {Baibhav}, {Barausse}, {Bonilla Rivera}, {Branchesi},
  {Branduardi-Raymont}, {Burdge}, {Chakraborty}, {Cuadra}, {Dage}, {Davis}, {de
  Mink}, {Decarli}, {Doneva}, {Escoffier}, {Gandhi}, {Haardt}, {Lousto},
  {Nissanke}, {Nordhaus}, {O'Shaughnessy}, {Portegies Zwart}, {Pound},
  {Schussler}, {Sergijenko}, {Spallicci}, {Vernieri}, \&
  {Vigna-G{\'o}mez}}]{2023LRR....26....2A}
{Amaro-Seoane}, P., {Andrews}, J., {Arca Sedda}, M., {et~al.} 2023, Living
  Reviews in Relativity, 26, 2

\bibitem[{{Amaro-Seoane} {et~al.}(2017){Amaro-Seoane}, {Audley}, {Babak},
  {Baker}, {Barausse}, {Bender}, {Berti}, {Binetruy}, {Born}, {Bortoluzzi},
  {Camp}, {Caprini}, {Cardoso}, {Colpi}, {Conklin}, {Cornish}, {Cutler},
  {Danzmann}, {Dolesi}, {Ferraioli}, {Ferroni}, {Fitzsimons}, {Gair}, {Gesa
  Bote}, {Giardini}, {Gibert}, {Grimani}, {Halloin}, {Heinzel}, {Hertog},
  {Hewitson}, {Holley-Bockelmann}, {Hollington}, {Hueller}, {Inchauspe},
  {Jetzer}, {Karnesis}, {Killow}, {Klein}, {Klipstein}, {Korsakova}, {Larson},
  {Livas}, {Lloro}, {Man}, {Mance}, {Martino}, {Mateos}, {McKenzie},
  {McWilliams}, {Miller}, {Mueller}, {Nardini}, {Nelemans}, {Nofrarias},
  {Petiteau}, {Pivato}, {Plagnol}, {Porter}, {Reiche}, {Robertson},
  {Robertson}, {Rossi}, {Russano}, {Schutz}, {Sesana}, {Shoemaker}, {Slutsky},
  {Sopuerta}, {Sumner}, {Tamanini}, {Thorpe}, {Troebs}, {Vallisneri},
  {Vecchio}, {Vetrugno}, {Vitale}, {Volonteri}, {Wanner}, {Ward}, {Wass},
  {Weber}, {Ziemer}, \& {Zweifel}}]{2017arXiv170200786A}
{Amaro-Seoane}, P., {Audley}, H., {Babak}, S., {et~al.} 2017, arXiv e-prints,
  arXiv:1702.00786

\bibitem[{{Aso} {et~al.}(2013){Aso}, {Michimura}, {Somiya}, {Ando}, {Miyakawa},
  {Sekiguchi}, {Tatsumi}, \& {Yamamoto}}]{2013PhRvD..88d3007A}
{Aso}, Y., {Michimura}, Y., {Somiya}, K., {et~al.} 2013, \prd, 88, 043007

\bibitem[{{Auclair} {et~al.}(2023){Auclair}, {Bacon}, {Baker}, {Barreiro},
  {Bartolo}, {Belgacem}, {Bellomo}, {Ben-Dayan}, {Bertacca}, {Besancon},
  {Blanco-Pillado}, {Blas}, {Boileau}, {Calcagni}, {Caldwell}, {Caprini},
  {Carbone}, {Chang}, {Chen}, {Christensen}, {Clesse}, {Comelli}, {Congedo},
  {Contaldi}, {Crisostomi}, {Croon}, {Cui}, {Cusin}, {Cutting}, {Dalang}, {De
  Luca}, {Pozzo}, {Desjacques}, {Dimastrogiovanni}, {Dorsch}, {Ezquiaga},
  {Fasiello}, {Figueroa}, {Flauger}, {Franciolini}, {Frusciante}, {Fumagalli},
  {Garc{\'\i}a-Bellido}, {Gould}, {Holz}, {Iacconi}, {Jain}, {Jenkins},
  {Jinno}, {Joana}, {Karnesis}, {Konstandin}, {Koyama}, {Kozaczuk},
  {Kuroyanagi}, {Laghi}, {Lewicki}, {Lombriser}, {Madge}, {Maggiore},
  {Malhotra}, {Mancarella}, {Mandic}, {Mangiagli}, {Matarrese}, {Mazumdar},
  {Mukherjee}, {Musco}, {Nardini}, {No}, {Papanikolaou}, {Peloso}, {Pieroni},
  {Pilo}, {Raccanelli}, {Renaux-Petel}, {Renzini}, {Ricciardone}, {Riotto},
  {Romano}, {Rollo}, {Pol}, {Morales}, {Sakellariadou}, {Saltas}, {Scalisi},
  {Schmitz}, {Schwaller}, {Sergijenko}, {Servant}, {Simakachorn}, {Sorbo},
  {Sousa}, {Speri}, {Steer}, {Tamanini}, {Tasinato}, {Torrado}, {Unal},
  {Vennin}, {Vernieri}, {Vernizzi}, {Volonteri}, {Wachter}, {Wands},
  {Witkowski}, {Zumalac{\'a}rregui}, {Annis}, {Ares}, {Avelino}, {Avgoustidis},
  {Barausse}, {Bonilla}, {Bonvin}, {Bosso}, {Calabrese},
  {{\c{c}}al{\i}{\c{s}}kan}, {Cembranos}, {Chala}, {Chernoff}, {Clough},
  {Criswell}, {Das}, {Silva}, {Dayal}, {Domcke}, {Durrer}, {Easther},
  {Escoffier}, {Ferrans}, {Fryer}, {Gair}, {Gordon}, {Hendry}, {Hindmarsh},
  {Hooper}, {Kajfasz}, {Kopp}, {Koushiappas}, {Kumar}, {Kunz}, {Lagos},
  {Lilley}, {Lizarraga}, {Lobo}, {Maleknejad}, {Martins}, {Meerburg}, {Meyer},
  {Mimoso}, {Nesseris}, {Nunes}, {Oikonomou}, {Orlando}, {{\"O}zsoy},
  {Pacucci}, {Palmese}, {Petiteau}, {Pinol}, {Zwart}, {Pratten}, {Prokopec},
  {Quenby}, {Rastgoo}, {Roest}, {Rummukainen}, {Schimd}, {Secroun}, {Sesana},
  {Sopuerta}, {Tereno}, {Tolley}, {Urrestilla}, {Vagenas}, {van de Vis}, {van
  de Weygaert}, {Wardell}, {Weir}, {White}, {{\'S}wie{\.Z}ewska}, {Zhdanov}, \&
  {LISA Cosmology Working Group}}]{2023LRR....26....5A}
{Auclair}, P., {Bacon}, D., {Baker}, T., {et~al.} 2023, Living Reviews in
  Relativity, 26, 5

\bibitem[{{Babak} {et~al.}(2023){Babak}, {Caprini}, {Figueroa}, {Karnesis},
  {Marcoccia}, {Nardini}, {Pieroni}, {Ricciardone}, {Sesana}, \&
  {Torrado}}]{2023JCAP...08..034B}
{Babak}, S., {Caprini}, C., {Figueroa}, D.~G., {et~al.} 2023, \jcap, 2023, 034

\bibitem[{{Badenes} {et~al.}(2018){Badenes}, {Mazzola}, {Thompson}, {Covey},
  {Freeman}, {Walker}, {Moe}, {Troup}, {Nidever}, {Allende Prieto}, {Andrews},
  {Barb{\'a}}, {Beers}, {Bovy}, {Carlberg}, {De Lee}, {Johnson}, {Lewis},
  {Majewski}, {Pinsonneault}, {Sobeck}, {Stassun}, {Stringfellow}, \&
  {Zasowski}}]{2018ApJ...854..147B}
{Badenes}, C., {Mazzola}, C., {Thompson}, T.~A., {et~al.} 2018, \apj, 854, 147

\bibitem[{{Baghi}(2022)}]{2022arXiv220412142B}
{Baghi}, Q. 2022, arXiv e-prints, arXiv:2204.12142

\bibitem[{{Bavera} {et~al.}(2022){Bavera}, {Franciolini}, {Cusin}, {Riotto},
  {Zevin}, \& {Fragos}}]{2022A&A...660A..26B}
{Bavera}, S.~S., {Franciolini}, G., {Cusin}, G., {et~al.} 2022, \aap, 660, A26

\bibitem[{{Bin{\'e}truy} {et~al.}(2012){Bin{\'e}truy}, {Boh{\'e}}, {Caprini},
  \& {Dufaux}}]{2012JCAP...06..027B}
{Bin{\'e}truy}, P., {Boh{\'e}}, A., {Caprini}, C., \& {Dufaux}, J.-F. 2012,
  \jcap, 2012, 027

\bibitem[{{Caprini} {et~al.}(2009){Caprini}, {Durrer}, \&
  {Servant}}]{2009JCAP...12..024C}
{Caprini}, C., {Durrer}, R., \& {Servant}, G. 2009, \jcap, 2009, 024

\bibitem[{Caprini \& Figueroa(2018)}]{Caprini_2018}
Caprini, C. \& Figueroa, D.~G. 2018, Classical and Quantum Gravity, 35, 163001

\bibitem[{{Chruslinska} {et~al.}(2018){Chruslinska}, {Belczynski}, {Klencki},
  \& {Benacquista}}]{2018MNRAS.474.2937C}
{Chruslinska}, M., {Belczynski}, K., {Klencki}, J., \& {Benacquista}, M. 2018,
  \mnras, 474, 2937

\bibitem[{{Chru{\'s}li{\'n}ska} {et~al.}(2020){Chru{\'s}li{\'n}ska},
  {Je{\v{r}}{\'a}bkov{\'a}}, {Nelemans}, \& {Yan}}]{2020A&A...636A..10C}
{Chru{\'s}li{\'n}ska}, M., {Je{\v{r}}{\'a}bkov{\'a}}, T., {Nelemans}, G., \&
  {Yan}, Z. 2020, \aap, 636, A10

\bibitem[{{Chru{\'s}li{\'n}ska} \& {Nelemans}(2019)}]{2019MNRAS.488.5300C}
{Chru{\'s}li{\'n}ska}, M. \& {Nelemans}, G. 2019, \mnras, 488, 5300

\bibitem[{{Chru{\'s}li{\'n}ska} {et~al.}(2021){Chru{\'s}li{\'n}ska},
  {Nelemans}, {Boco}, \& {Lapi}}]{2021MNRAS.508.4994C}
{Chru{\'s}li{\'n}ska}, M., {Nelemans}, G., {Boco}, L., \& {Lapi}, A. 2021,
  \mnras, 508, 4994

\bibitem[{{Cusin} {et~al.}(2018){Cusin}, {Dvorkin}, {Pitrou}, \&
  {Uzan}}]{2018PhRvL.120w1101C}
{Cusin}, G., {Dvorkin}, I., {Pitrou}, C., \& {Uzan}, J.-P. 2018, \prl, 120,
  231101

\bibitem[{{Dvorkin} {et~al.}(2016){Dvorkin}, {Vangioni}, {Silk}, {Uzan}, \&
  {Olive}}]{2016MNRAS.461.3877D}
{Dvorkin}, I., {Vangioni}, E., {Silk}, J., {Uzan}, J.-P., \& {Olive}, K.~A.
  2016, \mnras, 461, 3877

\bibitem[{{Evans} {et~al.}(1987){Evans}, {Iben}, \&
  {Smarr}}]{1987ApJ...323..129E}
{Evans}, C.~R., {Iben}, Icko, J., \& {Smarr}, L. 1987, \apj, 323, 129

\bibitem[{{Farmer} \& {Phinney}(2003)}]{2003MNRAS.346.1197F}
{Farmer}, A.~J. \& {Phinney}, E.~S. 2003, \mnras, 346, 1197

\bibitem[{{Giacobbo} \& {Mapelli}(2018)}]{2018MNRAS.480.2011G}
{Giacobbo}, N. \& {Mapelli}, M. 2018, \mnras, 480, 2011

\bibitem[{{Han}(1998)}]{1998MNRAS.296.1019H}
{Han}, Z. 1998, \mnras, 296, 1019

\bibitem[{{Hils} {et~al.}(1990){Hils}, {Bender}, \&
  {Webbink}}]{1990ApJ...360...75H}
{Hils}, D., {Bender}, P.~L., \& {Webbink}, R.~F. 1990, \apj, 360, 75

\bibitem[{{Karnesis} {et~al.}(2021){Karnesis}, {Babak}, {Pieroni}, {Cornish},
  \& {Littenberg}}]{2021PhRvD.104d3019K}
{Karnesis}, N., {Babak}, S., {Pieroni}, M., {Cornish}, N., \& {Littenberg}, T.
  2021, \prd, 104, 043019

\bibitem[{{Keim} {et~al.}(2023){Keim}, {Korol}, \&
  {Rossi}}]{2023MNRAS.521.1088K}
{Keim}, M.~A., {Korol}, V., \& {Rossi}, E.~M. 2023, \mnras, 521, 1088

\bibitem[{{Korol} {et~al.}(2022){Korol}, {Hallakoun}, {Toonen}, \&
  {Karnesis}}]{2022MNRAS.511.5936K}
{Korol}, V., {Hallakoun}, N., {Toonen}, S., \& {Karnesis}, N. 2022, \mnras,
  511, 5936

\bibitem[{{Korol} {et~al.}(2020){Korol}, {Toonen}, {Klein}, {Belokurov},
  {Vincenzo}, {Buscicchio}, {Gerosa}, {Moore}, {Roebber}, {Rossi}, \&
  {Vecchio}}]{2020A&A...638A.153K}
{Korol}, V., {Toonen}, S., {Klein}, A., {et~al.} 2020, \aap, 638, A153

\bibitem[{{Kowalska-Leszczynska} {et~al.}(2015){Kowalska-Leszczynska},
  {Regimbau}, {Bulik}, {Dominik}, \& {Belczynski}}]{2015A&A...574A..58K}
{Kowalska-Leszczynska}, I., {Regimbau}, T., {Bulik}, T., {Dominik}, M., \&
  {Belczynski}, K. 2015, \aap, 574, A58

\bibitem[{{Kroupa}(2001)}]{2001MNRAS.322..231K}
{Kroupa}, P. 2001, \mnras, 322, 231

\bibitem[{{Lamberts} {et~al.}(2019){Lamberts}, {Blunt}, {Littenberg},
  {Garrison-Kimmel}, {Kupfer}, \& {Sanderson}}]{2019MNRAS.490.5888L}
{Lamberts}, A., {Blunt}, S., {Littenberg}, T.~B., {et~al.} 2019, \mnras, 490,
  5888

\bibitem[{{Lehoucq} {et~al.}(2023){Lehoucq}, {Dvorkin}, {Srinivasan},
  {Pellouin}, \& {Lamberts}}]{2023MNRAS.526.4378L}
{Lehoucq}, L., {Dvorkin}, I., {Srinivasan}, R., {Pellouin}, C., \& {Lamberts},
  A. 2023, \mnras, 526, 4378

\bibitem[{{Li} {et~al.}(2023){Li}, {Chen}, {Ge}, {Chen}, \&
  {Han}}]{2023A&A...669A..82L}
{Li}, Z., {Chen}, X., {Ge}, H., {Chen}, H.-L., \& {Han}, Z. 2023, \aap, 669,
  A82

\bibitem[{{LIGO Scientific Collaboration} {et~al.}(2015){LIGO Scientific
  Collaboration}, {Aasi}, {Abbott}, {Abbott}, {Abbott}, {Abernathy}, {Ackley},
  {Adams}, {Adams}, {Addesso}, {Adhikari}, {Adya}, {Affeldt}, {Aggarwal},
  {Aguiar}, {Ain}, {Ajith}, {Alemic}, {Allen}, {Amariutei}, {Anderson},
  {Anderson}, {Arai}, {Araya}, {Arceneaux}, {Areeda}, {Ashton}, {Ast}, {Aston},
  {Aufmuth}, {Aulbert}, {Aylott}, {Babak}, {Baker}, {Ballmer}, {Barayoga},
  {Barbet}, {Barclay}, {Barish}, {Barker}, {Barr}, {Barsotti}, {Bartlett},
  {Barton}, {Bartos}, {Bassiri}, {Batch}, {Baune}, {Behnke}, {Bell}, {Bell},
  {Benacquista}, {Bergman}, {Bergmann}, {Berry}, {Betzwieser}, {Bhagwat},
  {Bhandare}, {Bilenko}, {Billingsley}, {Birch}, {Biscans}, {Biwer},
  {Blackburn}, {Blackburn}, {Blair}, {Blair}, {Bock}, {Bodiya}, {Bojtos},
  {Bond}, {Bork}, {Born}, {Bose}, {Brady}, {Braginsky}, {Brau}, {Bridges},
  {Brinkmann}, {Brooks}, {Brown}, {Brown}, {Brown}, {Buchman}, {Buikema},
  {Buonanno}, {Cadonati}, {Calder{\'o}n Bustillo}, {Camp}, {Cannon}, {Cao},
  {Capano}, {Caride}, {Caudill}, {Cavagli{\`a}}, {Cepeda}, {Chakraborty},
  {Chalermsongsak}, {Chamberlin}, {Chao}, {Charlton}, {Chen}, {Cho}, {Cho},
  {Chow}, {Christensen}, {Chu}, {Chung}, {Ciani}, {Clara}, {Clark}, {Collette},
  {Cominsky}, {Constancio}, {Cook}, {Corbitt}, {Cornish}, {Corsi}, {Costa},
  {Coughlin}, {Countryman}, {Couvares}, {Coward}, {Cowart}, {Coyne}, {Coyne},
  {Craig}, {Creighton}, {Creighton}, {Cripe}, {Crowder}, {Cumming},
  {Cunningham}, {Cutler}, {Dahl}, {Dal Canton}, {Damjanic}, {Danilishin},
  {Danzmann}, {Dartez}, {Dave}, {Daveloza}, {Davies}, {Daw}, {DeBra}, {Del
  Pozzo}, {Denker}, {Dent}, {Dergachev}, {DeRosa}, {DeSalvo}, {Dhurandhar},
  {D{\textasciiacute}{\i}az}, {Di Palma}, {Dojcinoski}, {Dominguez}, {Donovan},
  {Dooley}, {Doravari}, {Douglas}, {Downes}, {Driggers}, {Du}, {Dwyer},
  {Eberle}, {Edo}, {Edwards}, {Edwards}, {Effler}, {Eggenstein}, {Ehrens},
  {Eichholz}, {Eikenberry}, {Essick}, {Etzel}, {Evans}, {Evans},
  {Factourovich}, {Fairhurst}, {Fan}, {Fang}, {Farr}, {Farr}, {Favata}, {Fays},
  {Fehrmann}, {Fejer}, {Feldbaum}, {Ferreira}, {Fisher}, {Frei}, {Freise},
  {Frey}, {Fricke}, {Fritschel}, {Frolov}, {Fuentes-Tapia}, {Fulda}, {Fyffe},
  {Gair}, {Gaonkar}, {Gehrels}, {Gergely}, {Giaime}, {Giardina}, {Gleason},
  {Goetz}, {Goetz}, {Gondan}, {Gonz{\'a}lez}, {Gordon}, {Gorodetsky}, {Gossan},
  {Go{\ss}ler}, {Gr{\"a}f}, {Graff}, {Grant}, {Gras}, {Gray}, {Greenhalgh},
  {Gretarsson}, {Grote}, {Grunewald}, {Guido}, {Guo}, {Gushwa}, {Gustafson},
  {Gustafson}, {Hacker}, {Hall}, {Hammond}, {Hanke}, {Hanks}, {Hanna},
  {Hannam}, {Hanson}, {Hardwick}, {Harry}, {Harry}, {Hart}, {Hartman},
  {Haster}, {Haughian}, {Hee}, {Heintze}, {Heinzel}, {Hendry}, {Heng},
  {Heptonstall}, {Heurs}, {Hewitson}, {Hild}, {Hoak}, {Hodge}, {Hollitt},
  {Holt}, {Hopkins}, {Hosken}, {Hough}, {Houston}, {Howell}, {Hu}, {Huerta},
  {Hughey}, {Husa}, {Huttner}, {Huynh}, {Huynh-Dinh}, {Idrisy}, {Indik},
  {Ingram}, {Inta}, {Islas}, {Isler}, {Isogai}, {Iyer}, {Izumi}, {Jacobson},
  {Jang}, {Jawahar}, {Ji}, {Jim{\'e}nez-Forteza}, {Johnson}, {Jones}, {Jones},
  {Ju}, {Haris}, {Kalogera}, {Kandhasamy}, {Kang}, {Kanner}, {Katsavounidis},
  {Katzman}, {Kaufer}, {Kaufer}, {Kaur}, {Kawabe}, {Kawazoe}, {Keiser},
  {Keitel}, {Kelley}, {Kells}, {Keppel}, {Key}, {Khalaidovski}, {Khalili},
  {Khazanov}, {Kim}, {Kim}, {Kim}, {Kim}, {Kim}, {King}, {King}, {Kinzel},
  {Kissel}, {Klimenko}, {Kline}, {Koehlenbeck}, {Kokeyama}, {Kondrashov},
  {Korobko}, {Korth}, {Kozak}, {Kringel}, {Krishnan}, {Krueger}, {Kuehn},
  {Kumar}, {Kumar}, {Kuo}, {Landry}, {Lantz}, {Larson}, {Lasky}, {Lazzarini},
  {Lazzaro}, {Le}, {Leaci}, {Leavey}, {Lebigot}, {Lee}, {Lee}, {Lee}, {Leong},
  {Levin}, {Levine}, {Lewis}, {Li}, {Libbrecht}, {Libson}, {Lin}, {Littenberg},
  {Lockerbie}, {Lockett}, {Logue}, {Lombardi}, {Lormand}, {Lough}, {Lubinski},
  {L{\"u}ck}, {Lundgren}, {Lynch}, {Ma}, {Macarthur}, {MacDonald},
  {Machenschalk}, {MacInnis}, {Macleod}, {Maga{\~n}a-Sandoval}, {Magee},
  {Mageswaran}, {Maglione}, {Mailand}, {Mandel}, {Mandic}, {Mangano},
  {Mansell}, {M{\'a}rka}, {M{\'a}rka}, {Markosyan}, {Maros}, {Martin},
  {Martin}, {Martynov}, {Marx}, {Mason}, {Massinger}, {Matichard}, {Matone},
  {Mavalvala}, {Mazumder}, {Mazzolo}, {McCarthy}, {McClelland}, {McCormick},
  {McGuire}, {McIntyre}, {McIver}, {McLin}, {McWilliams}, {Meadors},
  {Meinders}, {Melatos}, {Mendell}, {Mercer}, {Meshkov}, {Messenger}, {Meyers},
  {Miao}, {Middleton}, {Mikhailov}, {Miller}, {Miller}, {Millhouse}, {Ming},
  {Mirshekari}, {Mishra}, {Mitra}, {Mitrofanov}, {Mitselmakher}, {Mittleman},
  {Moe}, {Mohanty}, {Mohapatra}, {Moore}, {Moraru}, {Moreno}, {Morriss},
  {Mossavi}, {Mow-Lowry}, {Mueller}, {Mueller}, {Mukherjee}, {Mullavey},
  {Munch}, {Murphy}, {Murray}, {Mytidis}, {Nash}, {Nayak}, {Necula}, {Nedkova},
  {Newton}, {Nguyen}, {Nielsen}, {Nissanke}, {Nitz}, {Nolting}, {Normandin},
  {Nuttall}, {Ochsner}, {O'Dell}, {Oelker}, {Ogin}, {Oh}, {Oh}, {Ohme},
  {Oppermann}, {Oram}, {O'Reilly}, {Ortega}, {O'Shaughnessy}, {Osthelder},
  {Ott}, {Ottaway}, {Ottens}, {Overmier}, {Owen}, {Padilla}, {Pai}, {Pai},
  {Palashov}, {Pal-Singh}, {Pan}, {Pankow}, {Pannarale}, {Pant}, {Papa},
  {Paris}, {Patrick}, {Pedraza}, {Pekowsky}, {Pele}, {Penn}, {Perreca},
  {Phelps}, {Pierro}, {Pinto}, {Pitkin}, {Poeld}, {Post}, {Poteomkin},
  {Powell}, {Prasad}, {Predoi}, {Premachandra}, {Prestegard}, {Price},
  {Principe}, {Privitera}, {Prix}, {Prokhorov}, {Puncken}, {P{\"u}rrer}, {Qin},
  {Quetschke}, {Quintero}, {Quiroga}, {Quitzow-James}, {Raab}, {Rabeling},
  {Radkins}, {Raffai}, {Raja}, {Rajalakshmi}, {Rakhmanov}, {Ramirez},
  {Raymond}, {Reed}, {Reid}, {Reitze}, {Reula}, {Riles}, {Robertson}, {Robie},
  {Rollins}, {Roma}, {Romano}, {Romanov}, {Romie}, {Rowan}, {R{\"u}diger},
  {Ryan}, {Sachdev}, {Sadecki}, {Sadeghian}, {Saleem}, {Salemi}, {Sammut},
  {Sandberg}, {Sanders}, {Sannibale}, {Santiago-Prieto}, {Sathyaprakash},
  {Saulson}, {Savage}, {Sawadsky}, {Scheuer}, {Schilling}, {Schmidt},
  {Schnabel}, {Schofield}, {Schreiber}, {Schuette}, {Schutz}, {Scott}, {Scott},
  {Sellers}, {Sengupta}, {Sergeev}, {Serna}, {Sevigny}, {Shaddock}, {Shahriar},
  {Shaltev}, {Shao}, {Shapiro}, {Shawhan}, {Shoemaker}, {Sidery}, {Siemens},
  {Sigg}, {Silva}, {Simakov}, {Singer}, {Singer}, {Singh}, {Sintes},
  {Slagmolen}, {Smith}, {Smith}, {Smith}, {Smith-Lefebvre}, {Son}, {Sorazu},
  {Souradeep}, {Staley}, {Stebbins}, {Steinke}, {Steinlechner}, {Steinlechner},
  {Steinmeyer}, {Stephens}, {Steplewski}, {Stevenson}, {Stone}, {Strain},
  {Strigin}, {Sturani}, {Stuver}, {Summerscales}, {Sutton}, {Szczepanczyk},
  {Szeifert}, {Talukder}, {Tanner}, {T{\'a}pai}, {Tarabrin}, {Taracchini},
  {Taylor}, {Tellez}, {Theeg}, {Thirugnanasambandam}, {Thomas}, {Thomas},
  {Thorne}, {Thorne}, {Thrane}, {Tiwari}, {Tomlinson}, {Torres}, {Torrie},
  {Traylor}, {Tse}, {Tshilumba}, {Ugolini}, {Unnikrishnan}, {Urban}, {Usman},
  {Vahlbruch}, {Vajente}, {Valdes}, {Vallisneri}, {van Veggel}, {Vass},
  {Vaulin}, {Vecchio}, {Veitch}, {Veitch}, {Venkateswara}, {Vincent-Finley},
  {Vitale}, {Vo}, {Vorvick}, {Vousden}, {Vyatchanin}, {Wade}, {Wade}, {Wade},
  {Walker}, {Wallace}, {Walsh}, {Wang}, {Wang}, {Wang}, {Ward}, {Warner},
  {Was}, {Weaver}, {Weinert}, {Weinstein}, {Weiss}, {Welborn}, {Wen},
  {Wessels}, {Westphal}, {Wette}, {Whelan}, {Whitcomb}, {White}, {Whiting},
  {Wilkinson}, {Williams}, {Williams}, {Williamson}, {Willis}, {Willke},
  {Wimmer}, {Winkler}, {Wipf}, {Wittel}, {Woan}, {Worden}, {Xie}, {Yablon},
  {Yakushin}, {Yam}, {Yamamoto}, {Yancey}, {Yang}, {Zanolin}, {Zhang}, {Zhang},
  {Zhang}, {Zhang}, {Zhao}, {Zhou}, {Zhu}, {Zucker}, {Zuraw}, \&
  {Zweizig}}]{2015CQGra..32g4001L}
{LIGO Scientific Collaboration}, {Aasi}, J., {Abbott}, B.~P., {et~al.} 2015,
  Classical and Quantum Gravity, 32, 074001

\bibitem[{{Lipunov} {et~al.}(1987){Lipunov}, {Postnov}, \&
  {Prokhorov}}]{1987A&A...176L...1L}
{Lipunov}, V.~M., {Postnov}, K.~A., \& {Prokhorov}, M.~E. 1987, \aap, 176, L1

\bibitem[{{Luo} {et~al.}(2016){Luo}, {Chen}, {Duan}, {Gong}, {Hu}, {Ji}, {Liu},
  {Mei}, {Milyukov}, {Sazhin}, {Shao}, {Toth}, {Tu}, {Wang}, {Wang}, {Yeh},
  {Zhan}, {Zhang}, {Zharov}, \& {Zhou}}]{2016CQGra..33c5010L}
{Luo}, J., {Chen}, L.-S., {Duan}, H.-Z., {et~al.} 2016, Classical and Quantum
  Gravity, 33, 035010

\bibitem[{{Luo} {et~al.}(2021){Luo}, {Wang}, {Wu}, {Hu}, \&
  {Jin}}]{2021PTEP.2021eA108L}
{Luo}, Z., {Wang}, Y., {Wu}, Y., {Hu}, W., \& {Jin}, G. 2021, Progress of
  Theoretical and Experimental Physics, 2021, 05A108

\bibitem[{Madau \& Dickinson(2014)}]{madau_cosmic_2014}
Madau, P. \& Dickinson, M. 2014, Annual Review of Astronomy and Astrophysics,
  52, 415

\bibitem[{{Moe} \& {Di Stefano}(2017)}]{2017ApJS..230...15M}
{Moe}, M. \& {Di Stefano}, R. 2017, \apjs, 230, 15

\bibitem[{{Moe} {et~al.}(2019){Moe}, {Kratter}, \&
  {Badenes}}]{2019ApJ...875...61M}
{Moe}, M., {Kratter}, K.~M., \& {Badenes}, C. 2019, \apj, 875, 61

\bibitem[{{Neijssel} {et~al.}(2019){Neijssel}, {Vigna-G{\'o}mez}, {Stevenson},
  {Barrett}, {Gaebel}, {Broekgaarden}, {de Mink}, {Sz{\'e}csi}, {Vinciguerra},
  \& {Mandel}}]{2019MNRAS.490.3740N}
{Neijssel}, C.~J., {Vigna-G{\'o}mez}, A., {Stevenson}, S., {et~al.} 2019,
  \mnras, 490, 3740

\bibitem[{{Nelemans} \& {Tout}(2005)}]{2005MNRAS.356..753N}
{Nelemans}, G. \& {Tout}, C.~A. 2005, \mnras, 356, 753

\bibitem[{{Nelemans} {et~al.}(2001{\natexlab{a}}){Nelemans}, {Yungelson}, \&
  {Portegies Zwart}}]{2001A&A...375..890N}
{Nelemans}, G., {Yungelson}, L.~R., \& {Portegies Zwart}, S.~F.
  2001{\natexlab{a}}, \aap, 375, 890

\bibitem[{{Nelemans} {et~al.}(2001{\natexlab{b}}){Nelemans}, {Yungelson},
  {Portegies Zwart}, \& {Verbunt}}]{2001A&A...365..491N}
{Nelemans}, G., {Yungelson}, L.~R., {Portegies Zwart}, S.~F., \& {Verbunt}, F.
  2001{\natexlab{b}}, \aap, 365, 491

\bibitem[{{Nissanke} {et~al.}(2012){Nissanke}, {Vallisneri}, {Nelemans}, \&
  {Prince}}]{2012ApJ...758..131N}
{Nissanke}, S., {Vallisneri}, M., {Nelemans}, G., \& {Prince}, T.~A. 2012,
  \apj, 758, 131

\bibitem[{{Phinney}(2001)}]{2001astro.ph..8028P}
{Phinney}, E.~S. 2001, arXiv:astro-ph/0108028

\bibitem[{{Portegies Zwart} \& {Verbunt}(1996)}]{1996A&A...309..179P}
{Portegies Zwart}, S.~F. \& {Verbunt}, F. 1996, \aap, 309, 179

\bibitem[{{Ruiter} {et~al.}(2010){Ruiter}, {Belczynski}, {Benacquista},
  {Larson}, \& {Williams}}]{2010ApJ...717.1006R}
{Ruiter}, A.~J., {Belczynski}, K., {Benacquista}, M., {Larson}, S.~L., \&
  {Williams}, G. 2010, \apj, 717, 1006

\bibitem[{{Schneider} {et~al.}(2001){Schneider}, {Ferrari}, {Matarrese}, \&
  {Portegies Zwart}}]{2001MNRAS.324..797S}
{Schneider}, R., {Ferrari}, V., {Matarrese}, S., \& {Portegies Zwart}, S.~F.
  2001, \mnras, 324, 797

\bibitem[{{Staelens} \& {Nelemans}(2024)}]{2024A&A...683A.139S}
{Staelens}, S. \& {Nelemans}, G. 2024, \aap, 683, A139

\bibitem[{{Temmink} {et~al.}(2023){Temmink}, {Pols}, {Justham}, {Istrate}, \&
  {Toonen}}]{2023A&A...669A..45T}
{Temmink}, K.~D., {Pols}, O.~R., {Justham}, S., {Istrate}, A.~G., \& {Toonen},
  S. 2023, \aap, 669, A45

\bibitem[{{Thrane} \& {Romano}(2013)}]{2013PhRvD..88l4032T}
{Thrane}, E. \& {Romano}, J.~D. 2013, \prd, 88, 124032

\bibitem[{{Toonen} {et~al.}(2012){Toonen}, {Nelemans}, \& {Portegies
  Zwart}}]{2012A&A...546A..70T}
{Toonen}, S., {Nelemans}, G., \& {Portegies Zwart}, S. 2012, \aap, 546, A70

\bibitem[{{van Haaften} {et~al.}(2013){van Haaften}, {Nelemans}, {Voss},
  {Toonen}, {Portegies Zwart}, {Yungelson}, \& {van der
  Sluys}}]{2013A&A...552A..69V}
{van Haaften}, L.~M., {Nelemans}, G., {Voss}, R., {et~al.} 2013, \aap, 552, A69

\bibitem[{{Woods} {et~al.}(2012){Woods}, {Ivanova}, {van der Sluys}, \&
  {Chaichenets}}]{2012ApJ...744...12W}
{Woods}, T.~E., {Ivanova}, N., {van der Sluys}, M.~V., \& {Chaichenets}, S.
  2012, \apj, 744, 12

\bibitem[{{Yu} \& {Jeffery}(2013)}]{2013MNRAS.429.1602Y}
{Yu}, S. \& {Jeffery}, C.~S. 2013, \mnras, 429, 1602

\end{thebibliography}
%
\begin{appendix}
\section{Additional Figures} \label{appendix}


\g{In this appendix we provide additional figures that give more details of the different models.}

\begin{figure}[ht]
    \includegraphics[width=0.9\columnwidth]{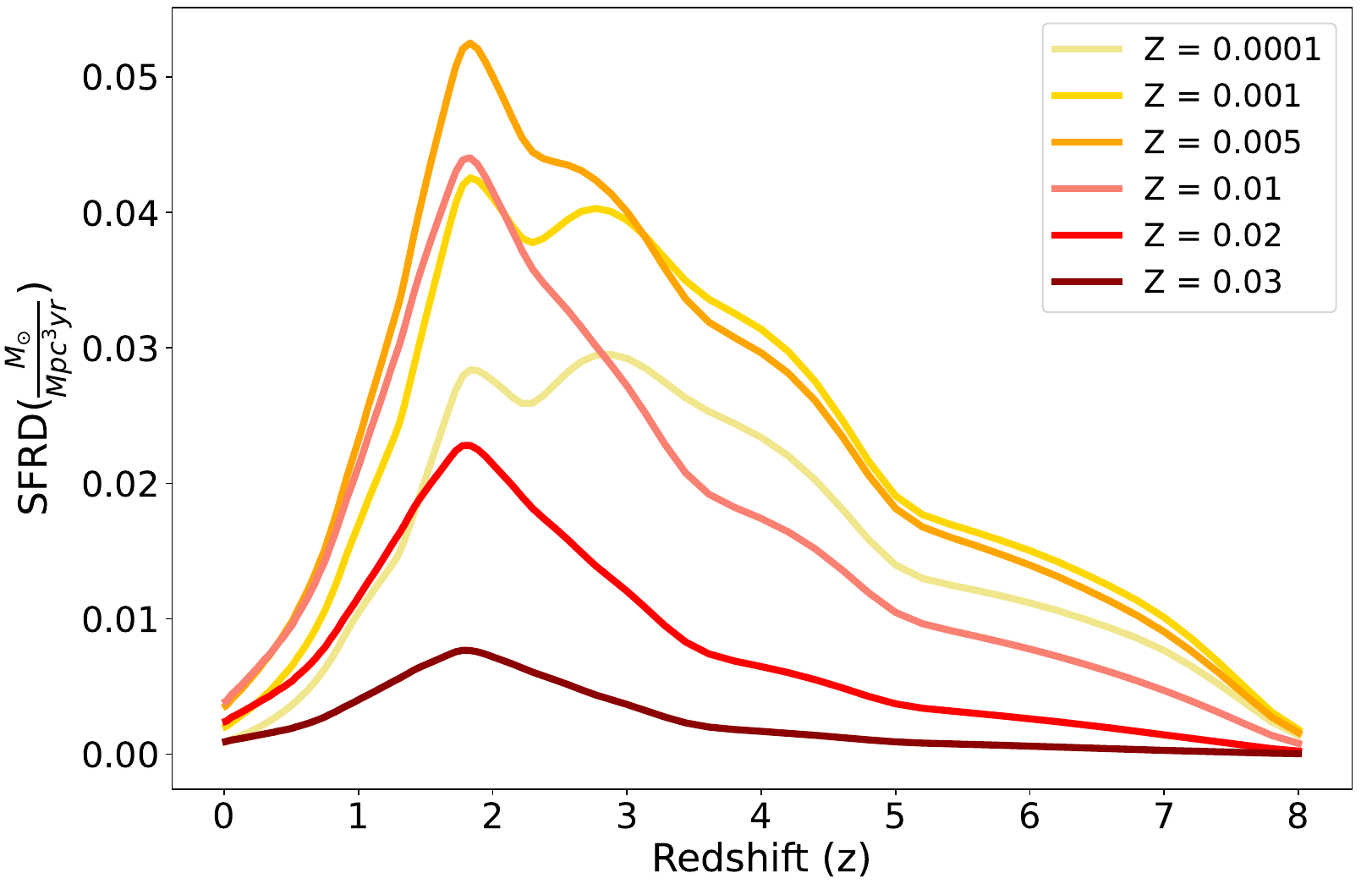}
    \caption{Total SFRD versus redshift (z), for the six metallicity bins for the LZ21 SFRD model.}
    \label{fig:SFRD vs z per bin LZ21}
\end{figure}

\begin{figure}[ht]
    \includegraphics[width=0.9\columnwidth]{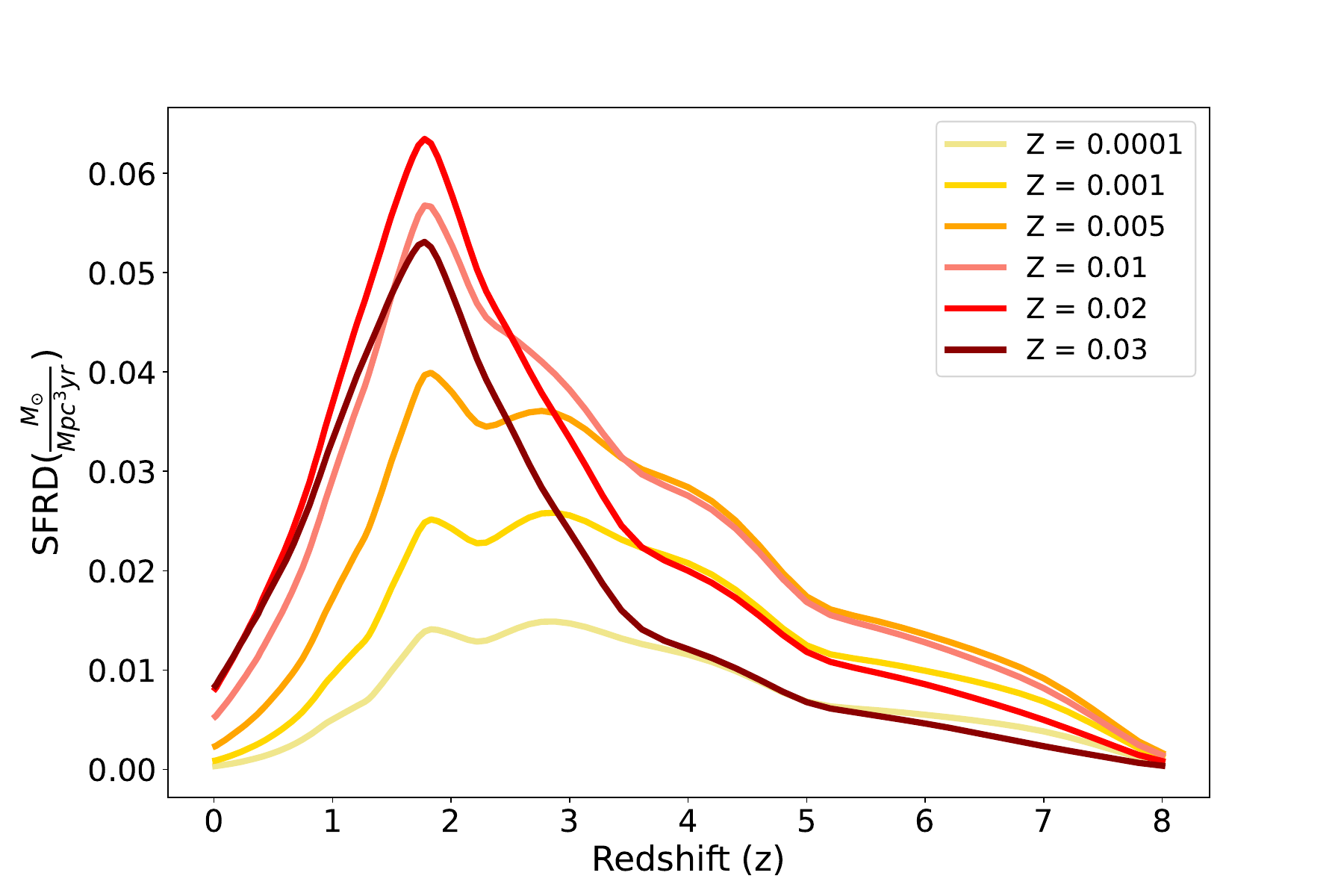}
    \caption{Total SFRD versus redshift (z), for the six metallicity bins for the HZ21 SFRD model.}
    \label{fig:SFRD vs z per bin HZ21}
\end{figure}

\g{Firstly, in the main paper we present the split of the SFRD over six different metallicity bins for the MZ19 model in Fig.~\ref{fig:SFRD vs z per bin MZ19}. Figs.~\ref{fig:SFRD vs z per bin LZ21} and Fig.~\ref{fig:SFRD vs z per bin HZ21} give the same for the LZ21 and the HZ21 SFRD models. The LZ19 and HZ19 SFRD model are left out, since these are very similar except for the normalisation due to the inclusion of starburst galaxies in the new models.}

\begin{figure*}
\sidecaption
\centering
\includegraphics[width=0.64\textwidth]{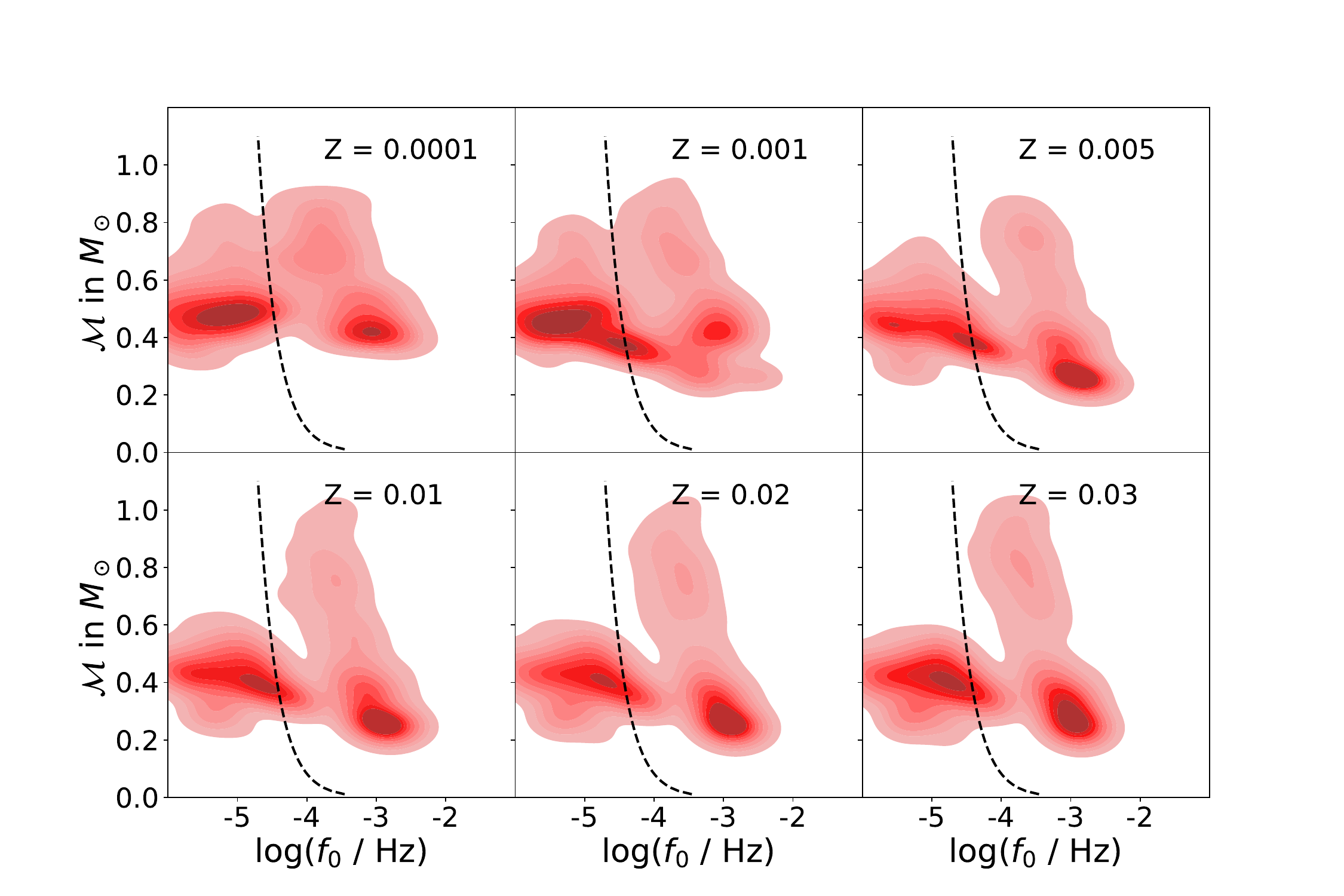}
    \caption{Density plot of the initial properties of the WD population in the case of a $\alpha\alpha$, $\alpha$ = 1 population synthesis model: chirp mass, $\mathcal{M}$, versus GW frequency at the time of formation. Each panel shows a different metallicity of the universe. The dashed lines indicate frequencies above which there is significant (10\%) frequency evolution in a Hubble time.}
    \label{fig:Density plots aa1}
\end{figure*}

\begin{figure*}
\sidecaption
    \centering
\includegraphics[width=0.64\textwidth]{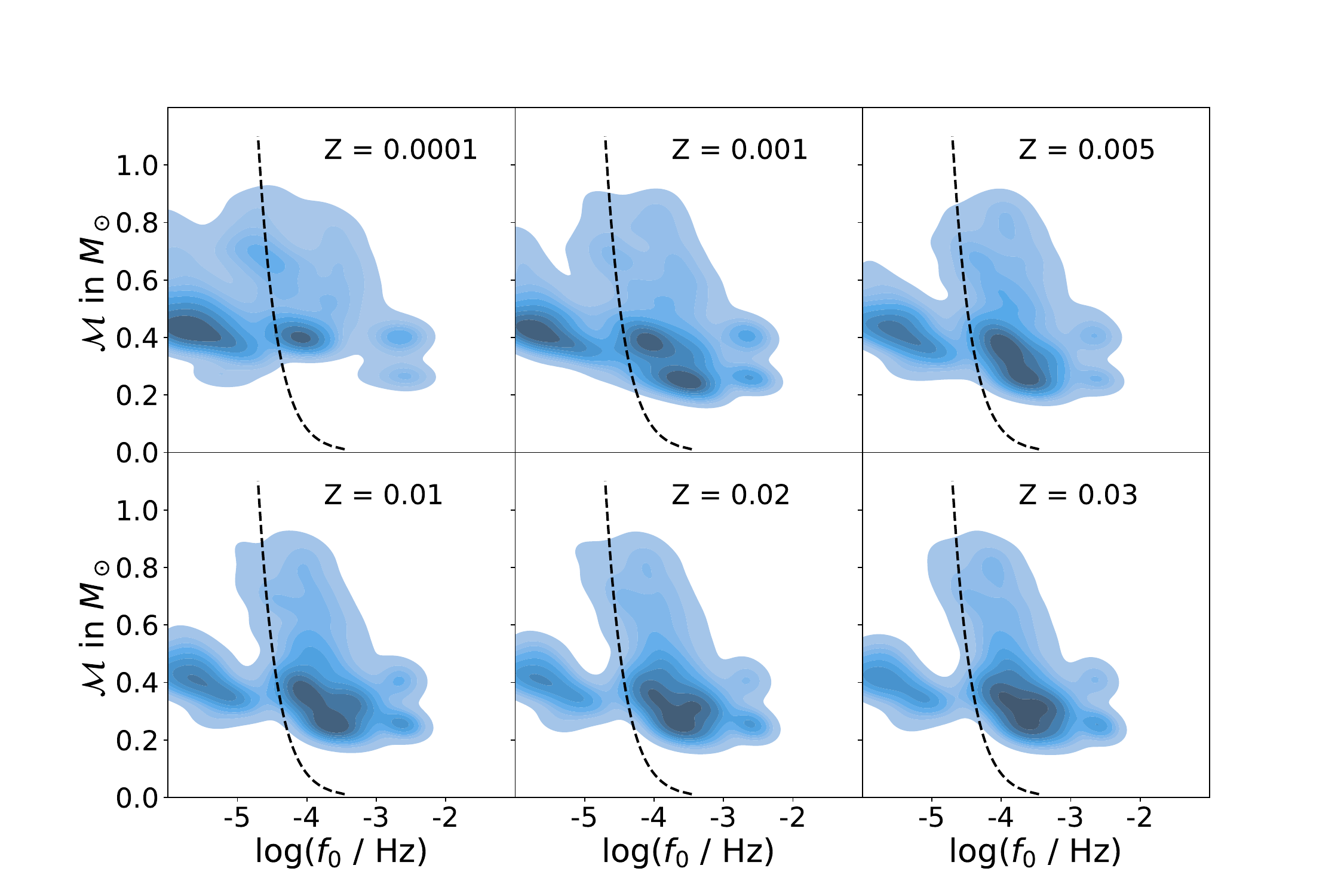}
    \caption{Density plot of the initial properties of the WD population in the case of a $\alpha\alpha$, $\alpha$ = 4 population synthesis model: chirp mass, $\mathcal{M}$, versus GW frequency at the time of formation. Each panel shows a different metallicity of the universe. The dashed lines indicate frequencies above which there is significant (10\%) frequency evolution in a Hubble time.}
    \label{fig:Density plots aa4}
\end{figure*}

\begin{figure*}
\sidecaption
    \centering
\includegraphics[width=0.64\textwidth]{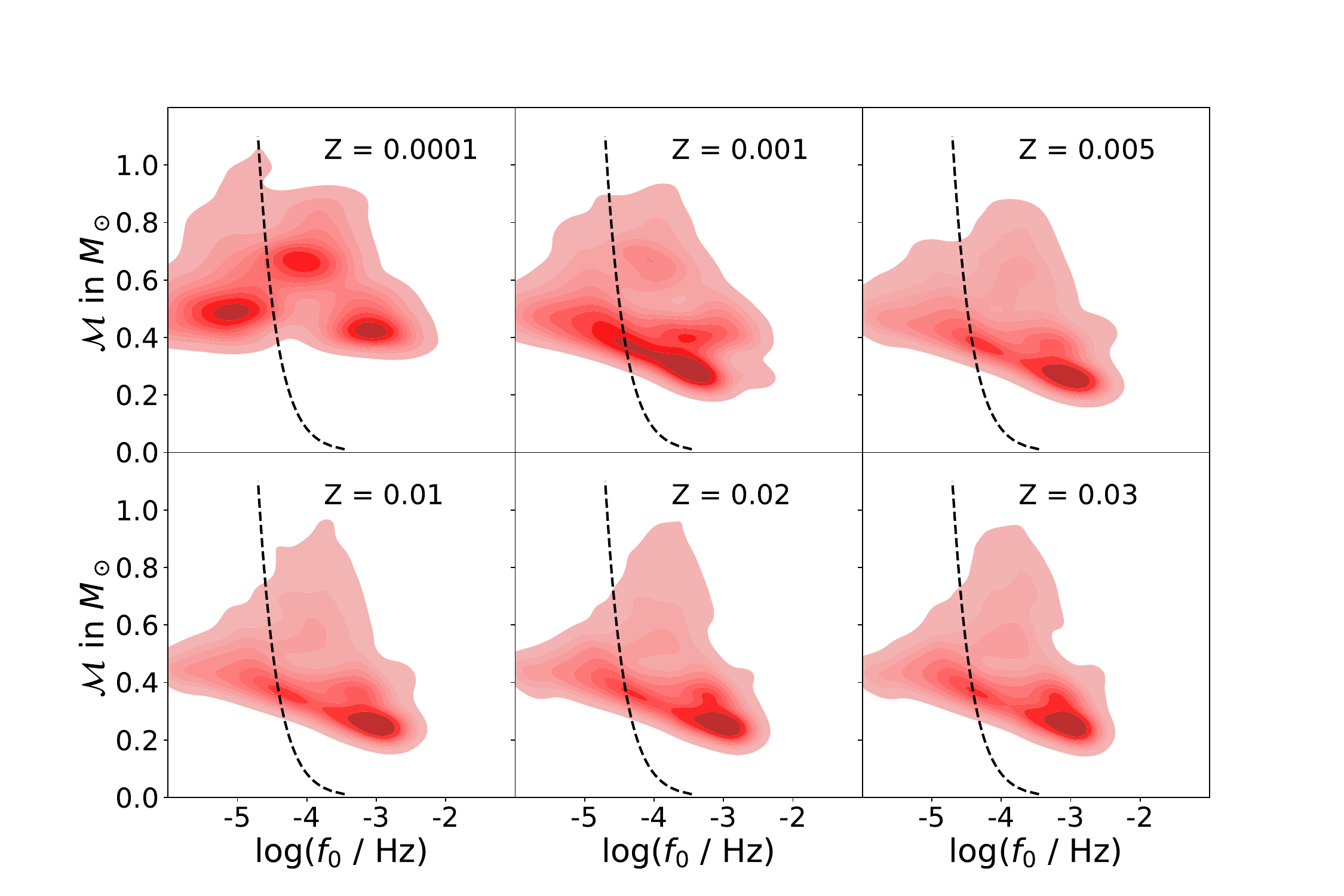}
    \caption{Density plot of the initial properties of the WD population in the case of a $\gamma\alpha$, $\alpha$ = 1 population synthesis model: chirp mass, $\mathcal{M}$, versus GW frequency at the time of formation. Each panel shows a different metallicity of the universe. The dashed lines indicate frequencies above which there is significant (10\%) frequency evolution in a Hubble time.}
    \label{fig:Density plots ga1}
\end{figure*}

\g{In the main text we presented the results for our default binary evolution model ($\gamma\alpha$, with $\alpha$ = 4).} \g{Figs.~\ref{fig:Density plots aa1} to \ref{fig:Density plots ga1} show the initial GW frequency and chirp mass for the different metallicities for the population models $\alpha\alpha$, with $\alpha$ = 1, $\alpha\alpha$, with $\alpha$ = 4 and $\gamma\alpha$, with $\alpha$ = 1, respectively. As for the standard model, most metallicities show very similar structure with gradual changes, mostly towards lower GW frequency and higher chirp mass for lower metallicity, with the lowest metallicity model often being quite different. As for the standard model, this is mostly due to the fact that stars are more compact at low metallicity.}

\begin{figure}[h]
    \includegraphics[width=0.93\columnwidth]{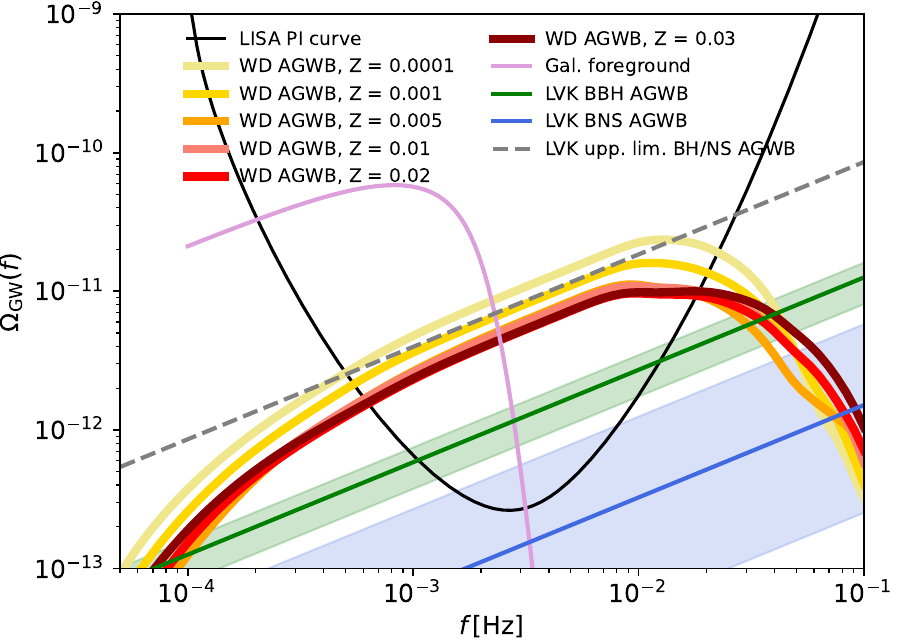}
    \caption{As in Figure \ref{fig:Omega ga4 SFH madau&dickinson}. The population synthesis model used is $\alpha\alpha$, $\alpha$ = 1 and the SFRD used is that of \cite{madau_cosmic_2014}.}
    \label{fig:Omega aa1 SFH madau&dickinson}
\end{figure}

\begin{figure}[h]
    \includegraphics[width=0.93\columnwidth]{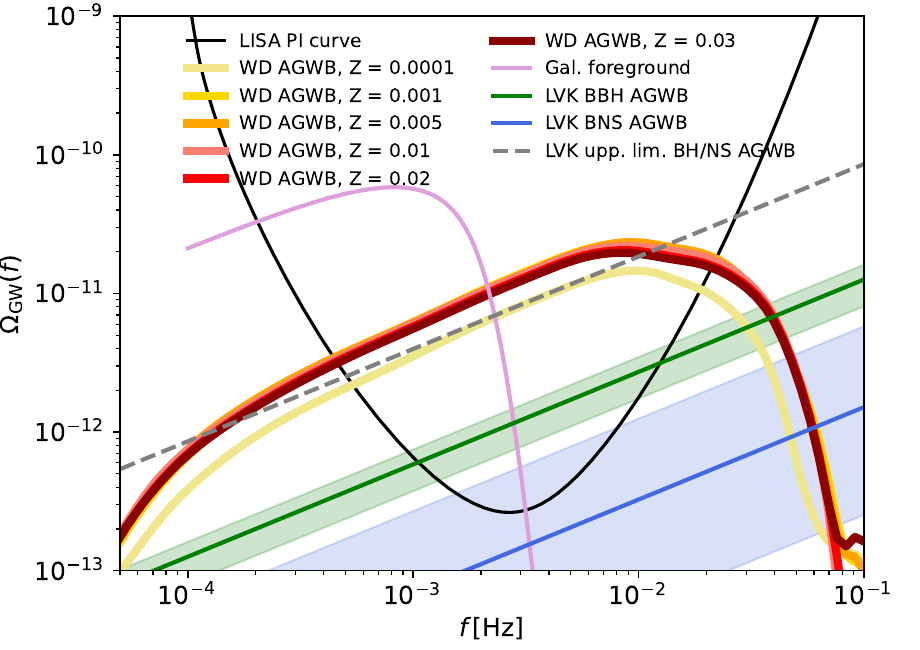}
    \caption{As in Figure \ref{fig:Omega ga4 SFH madau&dickinson}. The population synthesis model used is $\alpha\alpha$, $\alpha$ = 4 and the SFRD used is that of \cite{madau_cosmic_2014}.}
    \label{fig:Omega aa4 SFH madau&dickinson}
\end{figure}

\begin{figure}[h]
    \includegraphics[width=0.93\columnwidth]{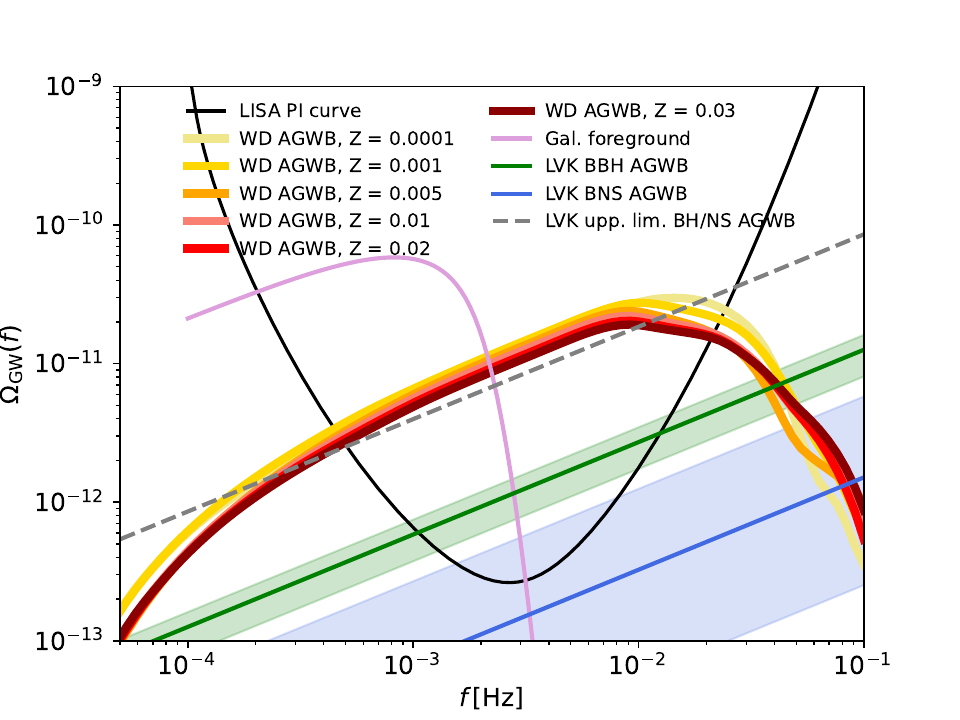}
    \caption{As in Figure \ref{fig:Omega ga4 SFH madau&dickinson}. The population synthesis model used is $\gamma\alpha$, $\alpha$ = 1 and the SFRD used is that of \cite{madau_cosmic_2014}.}
    \label{fig:Omega ga1 SFH madau&dickinson}
\end{figure}

\g{The resulting AGWB spectra for the hypothetical cases that all star formation would happen at a single metallicity are shown in  Fig.~\ref{fig:Omega aa1 SFH madau&dickinson} to \ref{fig:Omega ga1 SFH madau&dickinson}. These are the same as Fig.~\ref{fig:Omega ga4 SFH madau&dickinson}, except for the other population models. For the $\alpha\alpha$, with $\alpha$ = 1 model the shorter periods after the common envelope lead to an increasing number of sources with higher chirp mass towards lower metallicity, making the lower metallicity AGWBs higher than those of higher metallicity. This is contrary to in our standard model, but the overall AGWB is lower in these models. The $\alpha\alpha$, with $\alpha$ = 4 model shows similar behaviour to the standard model, with somewhat higher overall levels. Finally, in the $\gamma\alpha$, with $\alpha$ = 1 model, the increase in chrip mass towards lower metallicity is compensated with decrease in over all number, leading to very similar overall AGWB levels for all metallicities. }


\end{appendix}

\end{document}